\documentclass[useAMS,usenatbib]{mnras}

\usepackage{lscape}
\usepackage{graphicx}
\usepackage{multirow}
\usepackage{tabularx}  
\usepackage{txfonts}
\usepackage{caption}
\usepackage{subcaption}
\title[The effect of environment on the structure of disc galaxies]{The effect of environment on the structure of disc galaxies}
\author[Florian Pranger et al.]{Florian Pranger$^{1}$\thanks{E-mail:
florian@narit.or.th}, Ignacio Trujillo$^{2,3}$, Lee S. Kelvin$^{4}$, Mar\'ia Cebri\'{a}n$^{2,3}$\\
$^{1}$National Astronomical Research Institute of Thailand, 191 Huay Kaew Road, 50200 Chiang Mai, Thailand\\
$^{2}$Instituto de Astrof\'{i}sica de Canarias, c/V\'{i}a L\'{a}ctea s/n, E-38205-La Laguna, Tenerife, Spain\\
$^{3}$Departamento de Astrof\'{i}sica, Universidad de La Laguna, E-38205 La Laguna, Tenerife, Spain\\
$^{4}$Astrophysics Research Institute, Liverpool John Moores University, IC2, Liverpool Science Park, 146 Brownlow Hill, Liverpool L3 5RF, UK
}

\begin{document}


\pagerange{\pageref{firstpage}--\pageref{lastpage}} \pubyear{2016}

\maketitle

\label{firstpage}

\begin{abstract}

We study the influence of environment on the structure of disc galaxies, using \texttt{IMFIT} to measure the g- and r-band
structural parameters of the surface-brightness profiles for $\sim$700 low-redshift (z$<$0.063) cluster and field disc galaxies with
intermediate stellar mass (0.8 $\times$ 10$^{10}$ $M_{\odot}$ $<$ $M_{\star}$ $<$ 4 $\times$ 10$^{10}$ $M_{\odot}$) from the Sloan
Digital Sky Survey, DR7. Based on this measurement, we assign each galaxy to a surface-brightness profile type (Type I
$\equiv$ single-exponential, Type II $\equiv$ truncated, Type III $\equiv$ anti-truncated). In addition, we measure (g-r)
restframe colour for disc regions separated by the break radius. Cluster disc galaxies (at the same stellar mass) have
redder (g-r) colour by $\sim$0.2 mag than field galaxies. This reddening is slightly more pronounced outside the break radius.
Cluster disc galaxies also show larger global S\'ersic-indices and are more compact than field discs, both by $\sim$15\%. This change is connected to a flattening of the (outer) surface-brightness profile of Type I and - more significantly - of Type III galaxies by $\sim$8\% and $\sim$16\%, respectively, in the cluster environment compared to the field.
We find fractions of Type I, II and III of (6$\pm$2)\%, (66$\pm$4)\% and (29$\pm$4)\% in the field and (15$_{-4}^{+7}$)\%, (56$\pm$7)\% and (29$\pm$7)\% in the cluster environment, respectively. We suggest that the larger abundance of Type I galaxies in clusters (matched by a corresponding decrease in the Type II fraction) could be the
signature of a transition between Type II and Type I galaxies produced/enhanced by environment-driven mechanisms.  

\end{abstract}


\section{Introduction}

The last decade has witnessed the emergence of a new scheme to understand the structure of disc galaxies (e.g.
\citealt{erwin05a,pohlen06}). Following this picture, disc galaxies can be classified according to the different slopes they
present in their inner and outer regions. Galaxies which are best described by purely exponential surface-brightness
profiles \citep[e.g.][]{bland-hawthorn05, weiner01} are classified as Type I. Type II galaxies are those where the
surface-brightness profile is characterised by a broken exponential with a steep outer and a shallower inner region
\citep[e.g.][]{pohlen02}. Finally, Type III are disc galaxies the surface-brightness profiles of which are anti-truncated, i.e.  where the exponential decline of surface brightness is steeper in the inner
region and shallower at larger galactocentric distances \citep[e.g.][]{erwin05a}.\\
The emergence of Type II galaxies has been studied by a number of authors using numerical simulations (e.g. \citealt{li06, bournaud07, foyle08, roskar08, martinez-serrano09, sanchez-blazquez09}). To date, the general perception is that truncations that are not related to Lindblad-resonances (see e.g. \citealt{debattista06}) are the consequence of stellar migration and a radial star-formation threshold as the gas disc becomes thinner at large galactocentric radii. However, it is still being debated whether the other surface-brightness profile types also formed through internal mechanisms or are a consequence of external galaxy evolutionary processes. Theoretical work by e.g. \citet{yoshii89} and \citet{elmegreen06} shows that in principle, single-exponential and antitruncated discs can form ab initio, albeit under very specific conditions. However, it is worth noting that \citet{younger07} have shown that antitruncations can be the consequence of external processes like tidal interactions and minor galaxy mergers. Furthermore, in S0 galaxies, antitruncated profiles might also be related to an extended bulge component (e.g. \citealt{erwin05, maltby15}).\\
The first attempt to quantify the frequency of each profile type was done by \citet{pohlen06} (hereafter PT06). Using $\sim$90 nearby late-type (Sb-Sdm) spiral galaxies, they found that only $\sim$10\% are single-exponentials. The rest are $\sim$60\% Type II and $\sim$30\% anti-truncated galaxies. This study was conducted using disc galaxies of different environments. \citet{erwin08}, using a sample of 66 barred S0-Sb disc galaxies from different environments, found a distribution of 27\%, 42\% and 24\% of Type I, Type II and Type III profiles (with 6\% combinations of Type II and III). In a follow-up to this work (\citealt{gutierrez11}), the authors investigated a sample covering the full morphological range of disc galaxies (i.e. S0-Sm) and reported a correlation of morphology and surface-brightness profile type; Type I and Type III galaxies were found to be most frequent in early-type discs while the fraction of Type II profiles was found to increase with Hubble type, i.e. to be higher in late-type discs. Overall, they reported a distribution of 21\%, 50\% and 38\% of Type I, Type II and Type III profiles, with 8\% combined Type II and III galaxies that were counted twice.\\
A potential
change on the structural break properties of the disc galaxies is expected when comparing the field with the cluster
environment. There are many physical mechanisms in high density regions that should be particularly relevant for affecting
the outermost (weakly bound) zones of galaxies; tidal galaxy-galaxy interactions, tidal interactions between galaxies in high-density cluster regions and with the gravitational potential of the cluster (i.e. galaxy harassment) as well as hydrodynamical interactions between galaxies and the intra-cluster or intra-group medium such as ram-pressure stripping (see e.g. \citealt{toomre72, gunn72, moore96, lopes14, head14, hiemer14}).\\
The effect of the environment on the structure of galaxy discs is still a debated question and recent analyses have led to somewhat incongruent conclusions. For a sample of spiral galaxies, \citet{maltby15}
explored whether the frequency of each profile type changes in a cluster environment, finding  $\sim$10\% Type I, $\sim$50\% Type II and $\sim$40\% Type III galaxies in their cluster and field samples. The authors also investigated a sample of S0 cluster and field galaxies, and found $\sim$25\% Type I, $<$5\% Type II and $\sim$50\% Type III galaxies with $\sim$20\% of the profiles exhibiting general curvature and hence remaining unclassified. Comparing their field and cluster galaxies, the authors concluded that the stellar distribution in the outer regions of disc galaxies is not significantly affected by the galaxy environment. However, \citet{erwin12} reported that truncated (Type II) S0 galaxies are nonexistent in Virgo Cluster while they account for roughly one third of S0s in the local field. The difference in the cluster was found to be almost entirely compensated by Type I galaxies. The authors reasoned different mechanisms driving the structural evolution of galaxies in the cluster and field environment. Other works have also pointed to environment-mediated mechanisms and their effect on the structural
break properties of disc galaxies \citep[e.g.][]{roediger12,laine14,head14}.\\
\indent In this paper we want to investigate in detail whether the environment plays a role on the frequency of each profile
(break) type. To isolate as much as possible the effects produced by the cluster environment, we select our sample of
field and cluster galaxies to have the same redshifts and stellar mass ranges. We choose a narrow mass range since it has been shown that the structural parameters of disc galaxies change with mass (see PT06) and we want to minimise this effect in our analysis. We will show that the main effect of the cluster
environment is to redden by around $\sim$0.2 mag the (g-r) colour and to decrease the global size (as parameterised by the
effective radius $R_e$) of the discs by $\sim$15\%. These two global changes are accompanied by an increase (by a factor of
$\sim$2.5) in the fraction of Type I (pure exponential) disc galaxies in the cluster regions and an increase in the
(outer) scale lengths of Type I ($\sim$8\%) and Type III ($\sim$16\%) profiles.\\
In the subsequent section (Section \ref{sec:data}) we present a description of the data including a discussion of our selection criteria, the sample compilation and the background subtraction techniques. Section \ref{sec:fitting} describes the methods used for structural galaxy fitting and for colour determination. The results are presented in Section \ref{sec:results} and discussed in Section \ref{sec:discussion}. We summarise the results and conclusions in Section \ref{sec:summary}. Finally, we present prototypical profiles, further details of the analysis of the field sample and comprehensive data tables in an appendix to this paper.
\noindent Throughout this paper we assume $H_{0}$=70 km/s/Mpc, $\Omega_{m}$=0.3 and $\Omega_{\Lambda}$=0.7.

\begin{figure}
\vspace{-0.5cm}
\centering
\includegraphics[width=0.95\columnwidth]{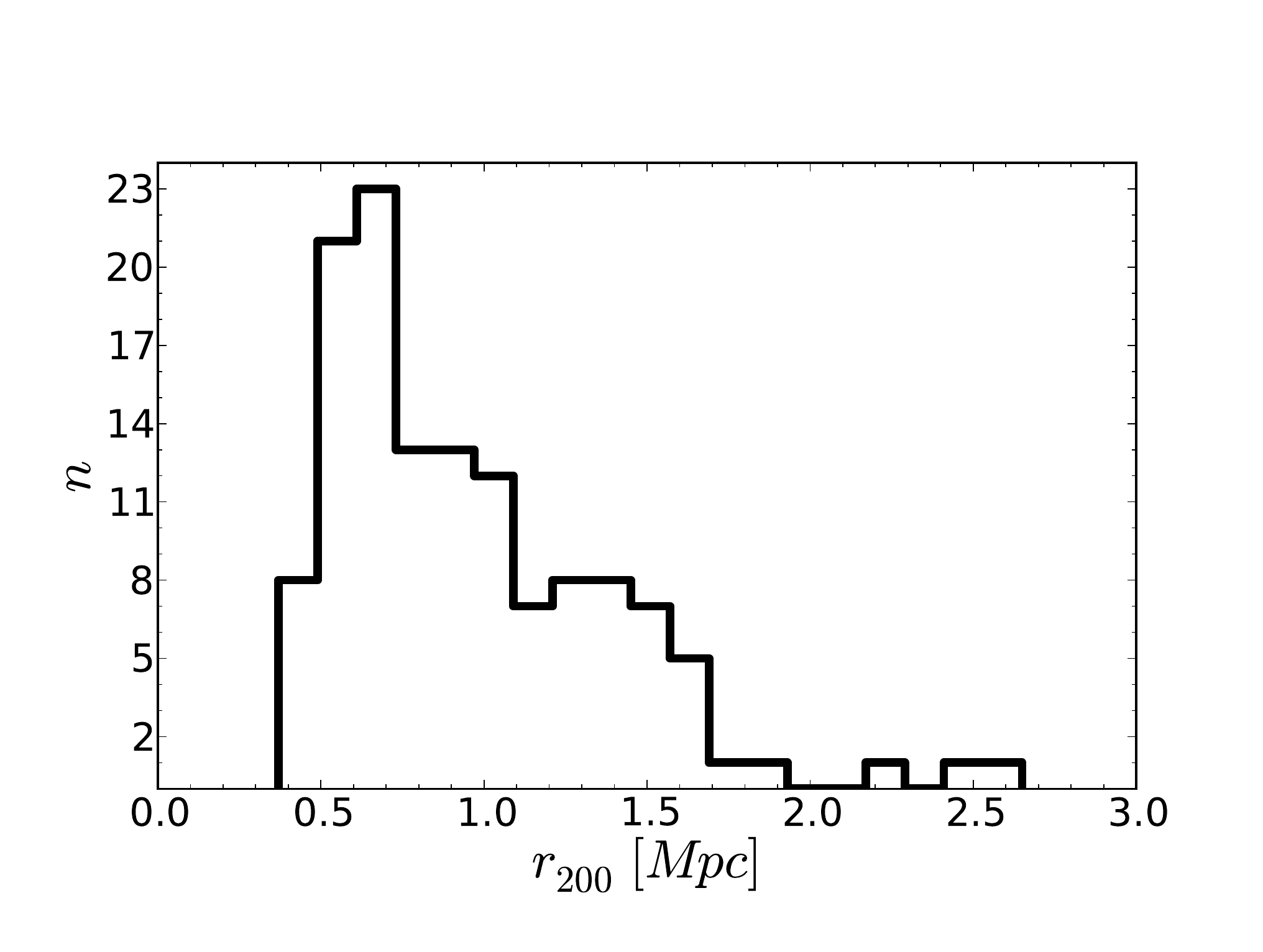}
\caption{Distribution of the $r_{200}$ values for all 130 host clusters.}
\label{fig:hist_r200}
\end{figure}

\begin{figure*}
\vspace{0.25cm}
\centering
\includegraphics[width=0.75\textwidth]{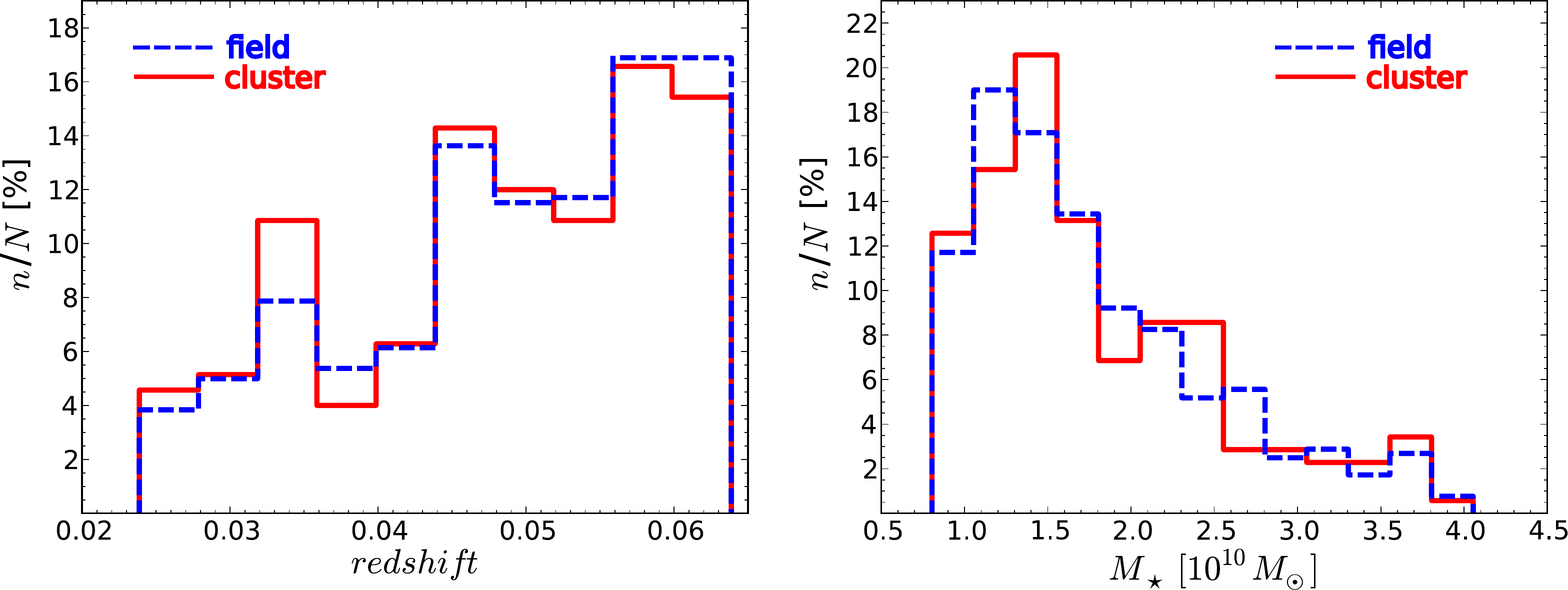}
\caption{\textit{Left panel:} normalised redshift distribution of the total cluster and total field sample. \textit{right panel:} normalised distribution of stellar mass for the total cluster and total field sample.}
\label{fig:1}
\end{figure*}

\section{The data}
\label{sec:data}

To conduct our work, we used the NYU Value-Added Galaxy Catalogue (NYU-VAGC, \citealt{blanton05}) based on the SDSS-DR7
\citep{abazajian09} as the basis for our parent sample. This catalogue provides spectroscopic redshifts and photometry, global S\'{e}rsic-indices and corresponding effective radii (see \citealt{blanton05a} for a detailed description of the morphological fitting technique) as well as stellar masses (\citealt{blanton07}) calculated on the basis of a \citet{chabrier03} IMF and a population synthesis model
from \citet{bruzual03}. Our parent sample is analogue to the one of \citet{cebrian14}. First, we selected only the galaxies contained within the region of the survey described by \citet{varela12}, in order to avoid problems with the borders of the survey. Then, we only considered the objects above the mass-completeness limit presented in \citet{cebrian14} (see Eq. 1 from that paper). This ensures that our sample is complete in stellar mass, avoiding biases due to the magnitude limit inherent to the SDSS catalogue. To assure that our initial sample of galaxies are predominantly disc-dominated and within a
narrow range of stellar masses, we took only objects with S\'{e}rsic-index n $<$ 2.5 and 0.8 $\times$ 10$^{10}$ $M_{\odot}$ $<$ $M_{\star}$ $<$ 4 $\times$ 10$^{10}$ $M_{\odot}$ The lower mass limit corresponds to our completeness limit while the higher mass limit was chosen to minimise the influence of stellar mass on the results of our study. \\
We also estimated, for all the galaxies, the 3D spatial (X, Y, Z) position within
the survey using the information provided by the R.A., Dec and redshift of each object. To do this, we used the following set of
equations provided by \citet{varela12}:

\begin{equation} \label{eq:cartesian}
\begin{array}{l}
X=D(z)\cos\delta\cos\alpha \\
Y=D(z)\cos\delta\sin\alpha \\
Z=D(z)\sin\delta \\
\end{array}
\end{equation}

\noindent with $\alpha$, $\delta$ and $D(z)$ being the equatorial coordinates and the comoving radial distance, respectively. The spatial distribution of the galaxies is used to explore the environment inhabited by the objects. 

\subsection{The cluster sample}

Following \citet{cebrian14}, we compiled a large sample of galaxy clusters in our explored volume using a number of catalogues: the
Abell catalogue \citep{abell89}, a catalogue extracted from SDSS-DR6 \citep{szabo11}, three catalogues from SDSS III \citep{einasto12, tempel12, wen12}; the GMBCG cluster catalogue \citep{hao10}, and the XMMi-SDSS
galaxy cluster survey \citep{takey11}. This is a total of 1877 galaxy clusters.\\
To build the sample of galaxies in clusters, we took only those galaxies from the parent sample that are located at a clustercentric distance
less than 1 Mpc to the nearest galaxy cluster centre. Since the goal of this work is to conduct a detailed analysis of the structural properties of
the disc of the galaxies, we selected the objects with the lowest redshifts. These selection criteria left us with 246 catalogue galaxies. To minimise
the influence of dust and to ensure the reliability of morphological information, we followed PT06 in selecting face-on to
intermediately inclined galaxies and we discarded close pairs and obvious galaxy mergers after a visual inspection. After doing this, 
we were left with 175 galaxies in 130 different clusters (25 clusters from \citealt{abell89}, 18 clusters from \citealt{szabo11}, 72
clusters from \citealt{tempel12} and 15 clusters from \citealt{wen12}). The sample of 175 galaxies will henceforth be referred to as the total cluster sample. The selection of a reasonable number of cluster galaxies finaly resulted in a redshift range of 0.021$<$z$<$0.063. All galaxies in the total cluster sample are listed in the appendix. The $r_{200}$\footnote{Values adopted from the cluster catalogues, except for Abell-clusters for which values were taken from the literature.}-distribution of the 130 host clusters is shown in Fig. \ref{fig:hist_r200}.

\subsection{The field sample}

In order to compare the properties of disc galaxies in the cluster environment to those of disc galaxies in the low-density to
intermediate-density field environment, we created a sample of field galaxies drawn from the same parent galaxy catalogue as the
initial cluster sample. In a first step, we confined the parent catalogue to galaxies with a 3D spatial distance greater than 3.5 Mpc
to the nearest cluster centre to ensure our objects are beyond the virial radius of the nearest cluster. In order to have a field sample with stellar mass and redshift distribution similar to
those distributions in the cluster sample, we generated random samples of 246 field galaxies (the same initial number as for the cluster galaxies),
and selected three samples closely resembling the initial cluster sample. As we did for the
cluster sample, we selected face-on to intermediately inclined galaxies and discarded close pairs and galaxy mergers after a by-eye
inspection. We were left with 172, 172 and 177 galaxies in the three field samples (no overlap). They will henceforth be referred to as field sample 1, field sample 2 and field sample 3. By joining the three field samples we constructed a large field sample of 521 galaxies which will henceforth be referred to as the total field sample. All galaxies in the total field sample are listed in the appendix. The redshift and stellar mass distributions of the total cluster and field samples are compared in Fig. \ref{fig:1}. The corresponding illustrations for the individual field samples are shown in Fig. \ref{fig:1a}.\\
We explored whether the global size of our field galaxies is representative of the general field population. With the redshift and
stellar mass limits of our cluster sample, we had 14868 field galaxies. Their median size (as parametrised by $R_{e}$) is larger than the median size of the cluster sample by
$\sim$13\%. Our field samples have median global sizes larger than the median size of our total cluster sample by $\sim$15\%. 
This difference in global size is in compliance with \citet{cebrian14}. 
All analyses presented in this paper were carried out on each of the initially selected samples (four in total). We use the three field samples to test for effects possibly introduced by random sampling and to estimate the robustness of the observed differences between the cluster and the field population.

\begin{figure*}
\vspace{0.25cm}
\centering
\includegraphics[width=\textwidth]{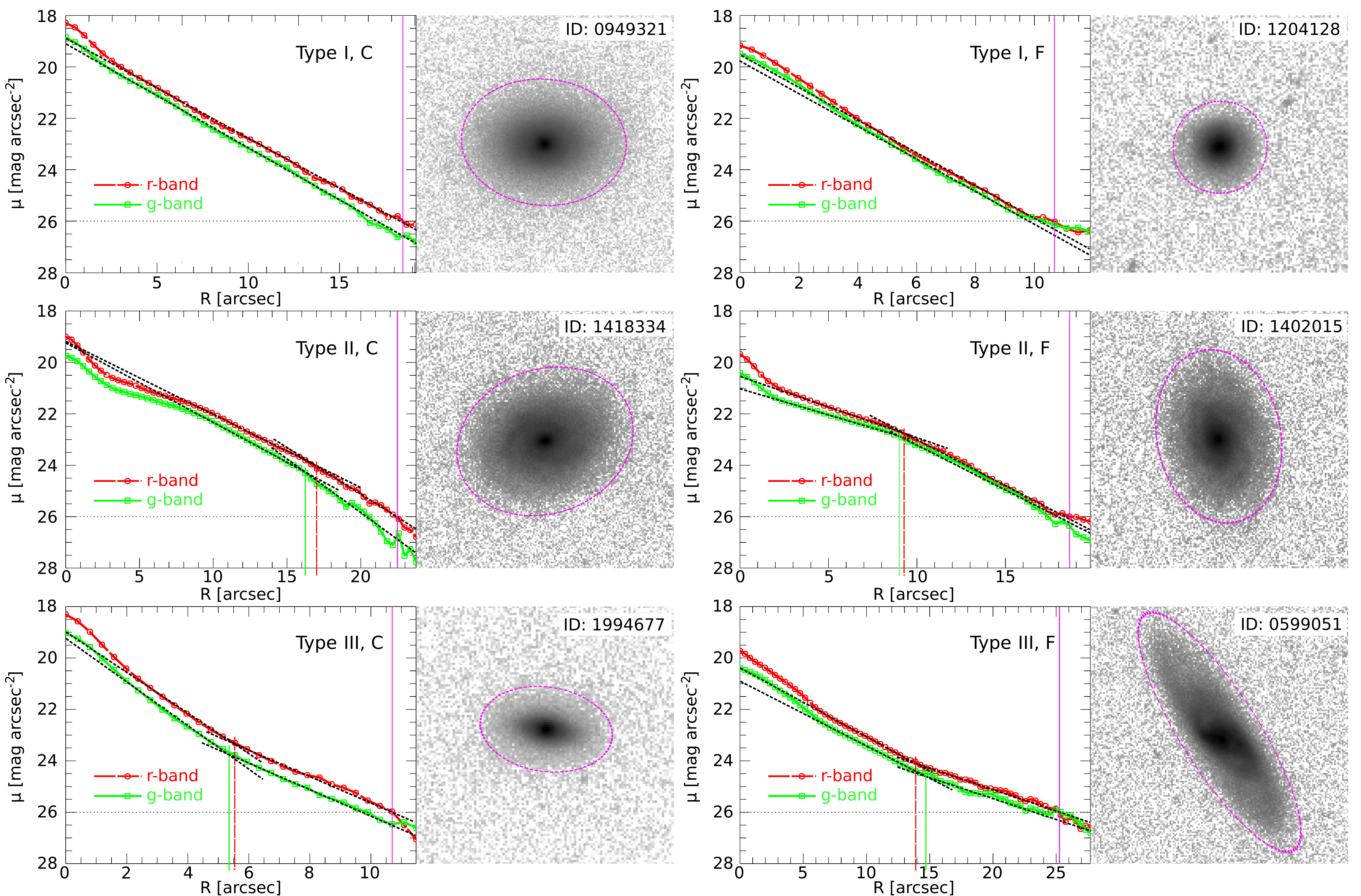}
\caption{Prototypical examples for each galaxy type in both environments. \textit{Left column, top to bottom:} Type I, Type II (double break) and Type III surface-brightness profiles and corresponding r-band cluster galaxy images. \textit{Right column, top to bottom:} Type I, Type II and Type III surface-brightness profiles and corresponding r-band field galaxy images. The over-plotted short-dashed lines in the profile plots indicate the exponential disc regions resulting from the fitting routine. The short vertical lines represent the break radii in the different bands. The horizontal dashed line indicates our conservative confidence threshold. For scale reference, the purple ellipses in the images mark the mean radius corresponding to this threshold, indicated by a solid vertical line in the profile plots. The galaxies' NYU-VAGC-ID is inserted in the images. 
}

\label{fig:2}
\end{figure*}

\subsection{Sky subtraction and limiting surface brightness}

As this work aims to explore the outer regions of disc galaxies to characterise their properties, it is necessary to have an
accurate sky subtraction and an estimation of the limiting surface brightness. Using the same SDSS imaging, PT06 showed that surface
brightness profiles in the g- and r-band can be reliably traced down to 27 mag/arcsec$^2$ (see also \citealt{trujillo16}). In order to
reach such accuracy, we followed the same strategy as PT06 to obtain an estimate of the sky from as close as possible to the galaxies under study. First, we
measured the mean sky after three 3-$\sigma$ clippings within 5 large rectangular sky boxes. Chosen sky boxes were placed as close as possible to the respective galaxy with the aim of including as little source flux as possible. This necessitates a variability in box size according to secondary source conditions unique to each pointing (typically in the range of 300-500 pixels on a side), with the total area fixed at 160k pixels per box. Any remaining extended sources within a sky box were masked out. For further details of this methodology, see PT06. As a second method, we investigated the radial
profile of the galaxy image pixel counts (with masked adjacent sources) in 120 angular directions, separated by 3 degrees. We controlled whether the sky value
measured with the first method was in compliance with the averaged value at which the radial profiles were visually found to flatten out. This
was true within $\pm$0.16 counts for all galaxies in our total field and cluster samples, both for g- and r-band data. For the photometric calibration we followed again PT06. This error in the sky determination corresponds to $\sim$16\% of the sky and to a surface brightness of $>$27 mag/arcsec$^{2}$. Since we do not attempt to extend our analyses beyond this value but limit ourselves to a conservative threshold of 26 mag/arcsec$^{2}$, this uncertainty has no effect on the results presented in this work.

\section{Structural galaxy fitting} 
\label{sec:fitting}
In order to characterise the structural (break) properties of our sample of 696 disc galaxies (175 cluster and 521
field), we used the surface-brightness 
fitting code \texttt{IMFIT} (\citealt{erwin15}). We applied this code to both, the g- and r-band data. The images provided to \texttt{IMFIT}
were masked and sky-subtracted. The masks were created using \texttt{SExtractor} (\citealt{bertin96}) in a hot and cold configuration
mode. To account for the effect of the Point Spread Function (PSF), we built a PSF for each SDSS image frame using \texttt{PSFExtractor}
\citep{bertin11} to estimate the FWHM of the PSF. Then, we used the image generator of the \texttt{IMFIT} package, \texttt{MAKEIMAGE}, to generate Moffat-PSF-images (SDSS standard $\beta$-values of 3.1 and 2.9 for the r- and g-band; see e.g. \citealt{trujillo01, erwin15} for more details) of
51$\times$51 pixels ($\sim$20\arcsec$\times$20\arcsec). The surface-brightness distributions of the galaxies in our sample were modelled
using a two-dimensional S\'ersic-bulge and a broken exponential (\citealt{erwin08}) for the disc component. These models were convolved with the
PSF.\\
\noindent
All 696 galaxies of the total cluster and field sample were successfully modelled by \texttt{IMFIT} in the g- and r-band. Each of the 1392 fits was set up individually, i.e. for each galaxy image we
created an individual configuration file with initial parameter values based on visual estimates after a by-eye inspection of the
azimuthally averaged 2D projected surface-brightness profile. The majority of the fits ran without any problems. For $\sim$30\% of the galaxies
it was necessary to further refine the initial fitting parameters. Approximately 40\% of these were to ensure that the innermost region of the galaxy (i.e. the bulge) was properly modelled by the
S\'{e}rsic part, while the disc was to be (dominantly) modelled by the broken exponential function to exclude potential degeneracies. For the remaining 60\%, the
profiles showed more than one break in the galaxy's disc region. In these cases, we manually forced \texttt{IMFIT} to model the
\textit{outermost} break above our limiting surface brightness. The profile types of the galaxies where the fit initially failed were Type II ($\sim$76\%) and III ($\sim$24\%).\\
For 56 galaxies (27 in the total cluster sample, 10, 10 and 9 in the field samples), the best fitting result was obtained for $h_{1}$=$h_{2}$
(i.e. the scale lengths in and outside the break have the same value). These galaxies were classified as Type I, following PT06. Another
199 galaxies (50 in the total cluster sample, 47, 50 and 52 in the field samples) showed an ``up-bending" or ``anti-truncated" profile, i.e. $h_{1}<h_{2}$, and were
classified as Type III. The remaining 441 galaxies (98 in the total cluster sample, 115, 112 and 116 in the field samples) exhibited a
"down-bending" profile, i.e. $h_{1}>h_{2}$, and thus were classified as Type II. The profile type number counts are the same in both bands. Depending on the analysed band, we notate the
structural parameters as follows: $h_{1,g}$, $h_{2,g}$, $R_{b,g}$ and $h_{1,r}$, $h_{2,r}$, $R_{b,r}$, where $R_{b,g}$ and $R_{b,r}$ denote the break radii. Profile type fractions and average parameter values are shown in Table \ref{table:results}. In Fig. \ref{fig:2} we show prototypical examples for the surface-brightness profiles and the fitted scale lengths for all three types in both investigated environments. \\
In PT06, the authors applied several sub-classifications to Type II and Type III galaxies, based on their Hubble-type (barred vs. non-barred
galaxies) and on the radial position of the break. Since the galaxy sample analysed for the present work are at a higher redshift
(0.021$\leq$z$\leq$0.063) than the sample analysed by PT06 (z$\leq$0.01), it was impossible to robustly determine the Hubble-type for all
galaxies in our study. Thus, we did not apply any further sub-classification for Type II and Type III galaxies. 

\subsection{Galaxy colour}
\label{sec:colour}
For all 696 galaxies in the combined total cluster and field sample, we measured (g-r) restframe colour, $(g-r)_{restframe}$, both
outside ($(g-r)_{o}$) and inside ($(g-r)_{i}$) the break radius $\overline{R_{b}}$=$(R_{b,g}+R_{b,r})/2$. The k-correction was conducted following \citet{chilingarian10}. In addition, to explore the effect of the bulge on the colour properties
of the inner regions of the galaxies, we repeated the above colour measurements but this time masking the inner $R_{c1}$=0.5$R_{e}$ (obtaining $(g-r)_{i,1}$) and
$R_{c2}$=0.75$R_{e}$ (obtaining $(g-r)_{i,2}$). For all investigated galaxies $R_b > R_{c1}$ and $R_b > R_{c2}$. The masking of the bulge regions results in a de-reddening of galaxy colours which is in general more pronounced in field galaxies and appears to be strongest in field Type II objects. This was to be expected, since Type II galaxies are predominantly late-type objects, which exhibit comparatively strong star-formation activity in their discs. The median colour values for the total cluster and total field sample are shown in Table \ref{table:results}, along with the other disc properties studied here.

\begin{table*}
\vspace{0.25cm}
\caption{Listing of the median values of the structural properties of the galaxies analysed in this work. The total cluster and total field sample
 are abbreviated by TCS and TFS, respectively. The ratios of the inner and outer scale lengths in the g- and r-band surface-brightness
profiles are listed as $ratio_{g}$ and $ratio_{r}$, respectively. Median colour values determined between $\overline{R_{b}}$ and
$R_{c1}$ are given as $(g-r)_{in,1}$, those determined between $\overline{R_{b}}$ and $R_{c2}$ as $(g-r)_{in,2}$ (see Section \ref{sec:colour}). Type I galaxies were not considered for the calculation of the median $(g-r)_{out}$ colour for the total samples. We also list the
number of galaxies in each sub-sample, N, and its percental equivalent (along with the corresponding \citealt{wilson27}-confidence intervals), as well as the median global effective radius $R_{e}$, median S\'{e}rsic-index $n$ and median stellar mass $M_{\star}$. All errors were estimated through 1000 1-$\sigma$ bootstrap
iterations. Field values with an asterisk indicate significant differences between the cluster and the field as estimated by a standard KS-test (P-value $<$0.05).}

\label{table:results}
\centering
\begin{tabular}{c|ccc|c||ccc|c}
\hline
\hline
\multicolumn{1}{c|}{} &
\multicolumn{4}{c||}{Cluster} & \multicolumn{4}{c}{Field}\\
\hline
Sample & Type I & Type II & Type III & TCS & Type I & Type II & Type III & TFS \\ 
\hline
N & 27 & 98 & 50 & 175 & 29 & 343 & 149 & 521 \\
\% & 15$_{-4}^{+7}$ & 56$_{-7}^{+7}$ & 29$_{-7}^{+7}$ & 100 & 6$_{-2}^{+2}$ & 66$_{-4}^{+4}$ & 29$_{-4}^{+4}$ & 100 \\
\hline
$R_{e}$ & 2.20$\pm$0.13 & 2.79$\pm$0.09 & 2.02$\pm$0.08 & 2.44$\pm$0.07 & 2.41$\pm$0.14 & 3.20$\pm$0.06* & 2.22$\pm$0.07 & 2.88$\pm$0.05* \\
$n$ & 2.07$\pm$0.04 & 1.83$\pm$0.04 & 2.17$\pm$0.03 & 1.99$\pm$0.03 & 1.60$\pm$0.06* & 1.63$\pm$0.02 & 2.00$\pm$0.03* & 1.73$\pm$0.02* \\
$M_{\star}$ [10$^{10}M_{\odot}$] & 1.52$\pm$0.10 & 1.53$\pm$0.05 & 1.89$\pm$0.11 & 1.63$\pm$0.05 & 1.55$\pm$0.14 & 1.60$\pm$0.03 & 1.55$\pm$0.06 & 1.59$\pm$0.03 \\
\hline
$(g-r)_{in}$ & 0.504$\pm$0.022 & 0.363$\pm$0.015 & 0.534$\pm$0.014 & 0.455$\pm$0.015 & 0.238$\pm$0.014* & 0.266$\pm$0.007* & 0.392$\pm$0.012* & 0.288$\pm$0.006* \\
$(g-r)_{in,1}$ & 0.504$\pm$0.022 & 0.334$\pm$0.015 & 0.506$\pm$0.014 & 0.434$\pm$0.011 & 0.222$\pm$0.012* & 0.235$\pm$0.007* & 0.362$\pm$0.012* & 0.265$\pm$0.006* \\
$(g-r)_{in,2}$ & 0.504$\pm$0.023 & 0.315$\pm$0.016 & 0.500$\pm$0.014 & 0.426$\pm$0.011 & 0.208$\pm$0.013* & 0.217$\pm$0.007* & 0.354$\pm$0.012* & 0.247$\pm$0.006* \\
$(g-r)_{out}$ & - & 0.300$\pm$0.020 & 0.482$\pm$0.026 & 0.363$\pm$0.015 & - & 0.201$\pm$0.010* & 0.288$\pm$0.015* & 0.231$\pm$0.008* \\
\hline
$h_{1,g}$ [kpc] & 1.55$\pm$0.08 & 2.75$\pm$0.14 & 1.38$\pm$0.06 & - & 1.47$\pm$0.05 & 2.85$\pm$0.07 & 1.37$\pm$0.04 & - \\
$h_{2,g}$ [kpc] & -"- & 1.37$\pm$0.05 & 2.18$\pm$0.11 & - & -"- & 1.46$\pm$0.03 & 1.83$\pm$0.05* & - \\
$ratio_{g}$ & 1 & 1.91$\pm$0.09 & 0.64$\pm$0.02 & - & 1 & 1.87$\pm$0.04 & 0.74$\pm$0.01* & - \\
$R_{b,g}$ [kpc] & - & 5.21$\pm$0.19 & 4.68$\pm$0.20 & - & - & 5.33$\pm$0.10 & 4.41$\pm$0.15 & - \\
$h_{1,r}$ [kpc] & 1.57$\pm$0.09 & 2.66$\pm$0.12 & 1.37$\pm$0.06 & - & 1.45$\pm$0.05 & 2.66$\pm$0.06 & 1.32$\pm$0.04 & - \\
$h_{2,r}$ [kpc] & -"- & 1.37$\pm$0.04 & 2.12$\pm$0.08 & - & -"- & 1.38$\pm$0.02 & 1.87$\pm$0.05* & - \\
$ratio_{r}$ & 1 & 1.71$\pm$0.10 & 0.63$\pm$0.02 & - & 1 & 1.69$\pm$0.04 & 0.71$\pm$0.01* & - \\
$R_{b,r}$ [kpc] & - & 5.72$\pm$0.19 & 4.74$\pm$0.20 & - & - & 5.53$\pm$0.10 & 5.22$\pm$0.15 & - \\
\hline
\hline
\end{tabular}
\vspace{0.5cm}
\end{table*}  

\begin{figure}
\vspace{-0.5cm}
\centering
\includegraphics[width=0.95\columnwidth]{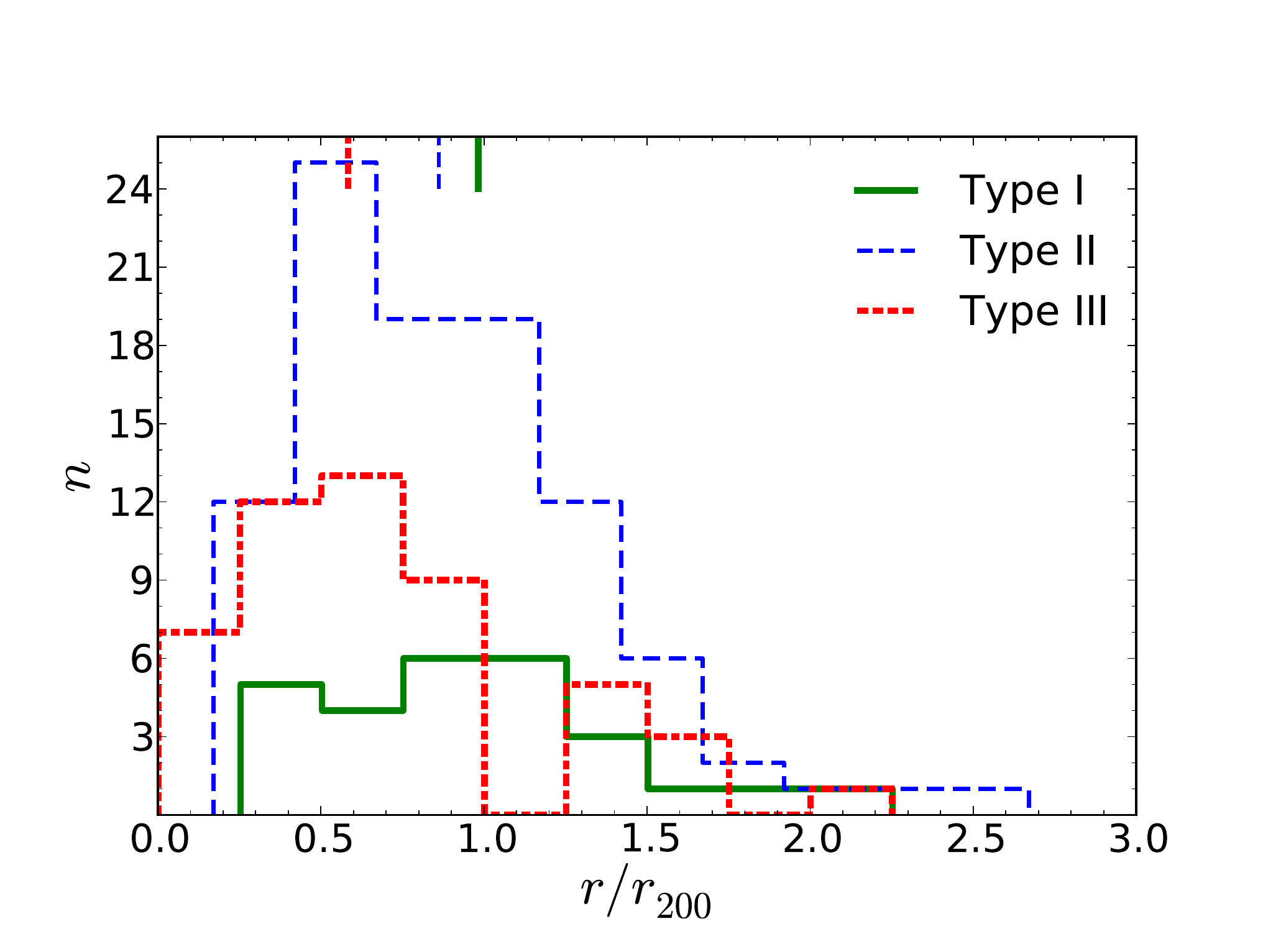}
\caption{Normalised radial distributions of the different profile types in the total cluster sample (green solid, blue dashed and red short-dashed lines for Type I, II and III, respectively). The short vertical lines indicate the corresponding median values.}
\label{fig:dist_norm}
\end{figure}

\section{Galaxy properties as a function of environment}
\label{sec:results}

\subsection{Global properties}

\begin{figure*}
\centering
\vspace{0.5cm}
\includegraphics[width=0.725\textwidth]{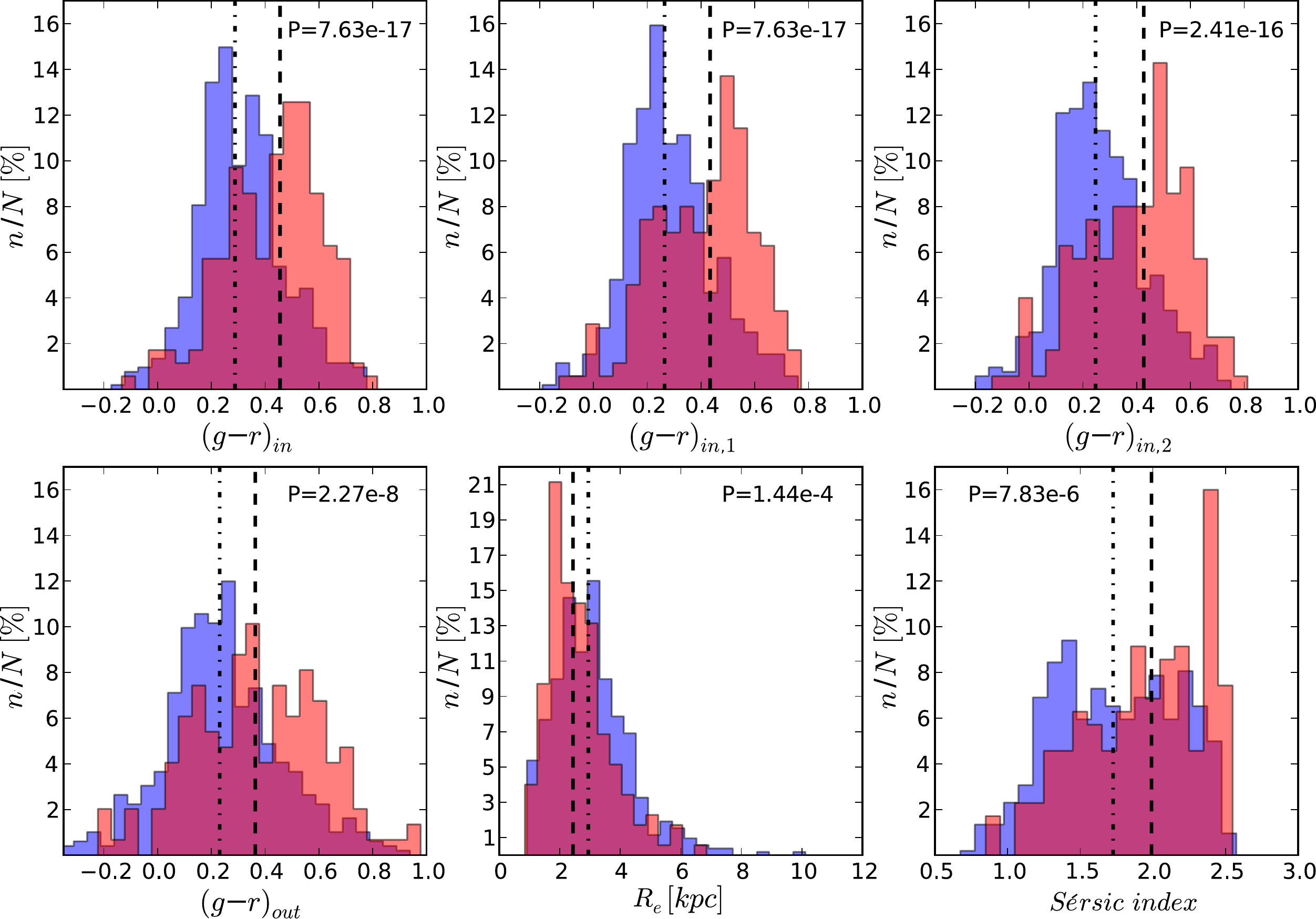}
\caption{Normalised distributions of colour, $R_{e}$ and S\'ersic-index for the total cluster and field samples (field galaxies in the background in blue colour, cluster galaxies in the foreground in light-red colour, overlapping regions of the histograms in purple). The vertical lines indicate the median values (dash-dotted for field and dashed for cluster galaxies). The KS-test P-value is inserted in each panel. Type I galaxies are not included in the bottom-left panel.}
\label{fig:firsthist}
\end{figure*}

\begin{figure*}
\centering
\includegraphics[width=0.95\textwidth]{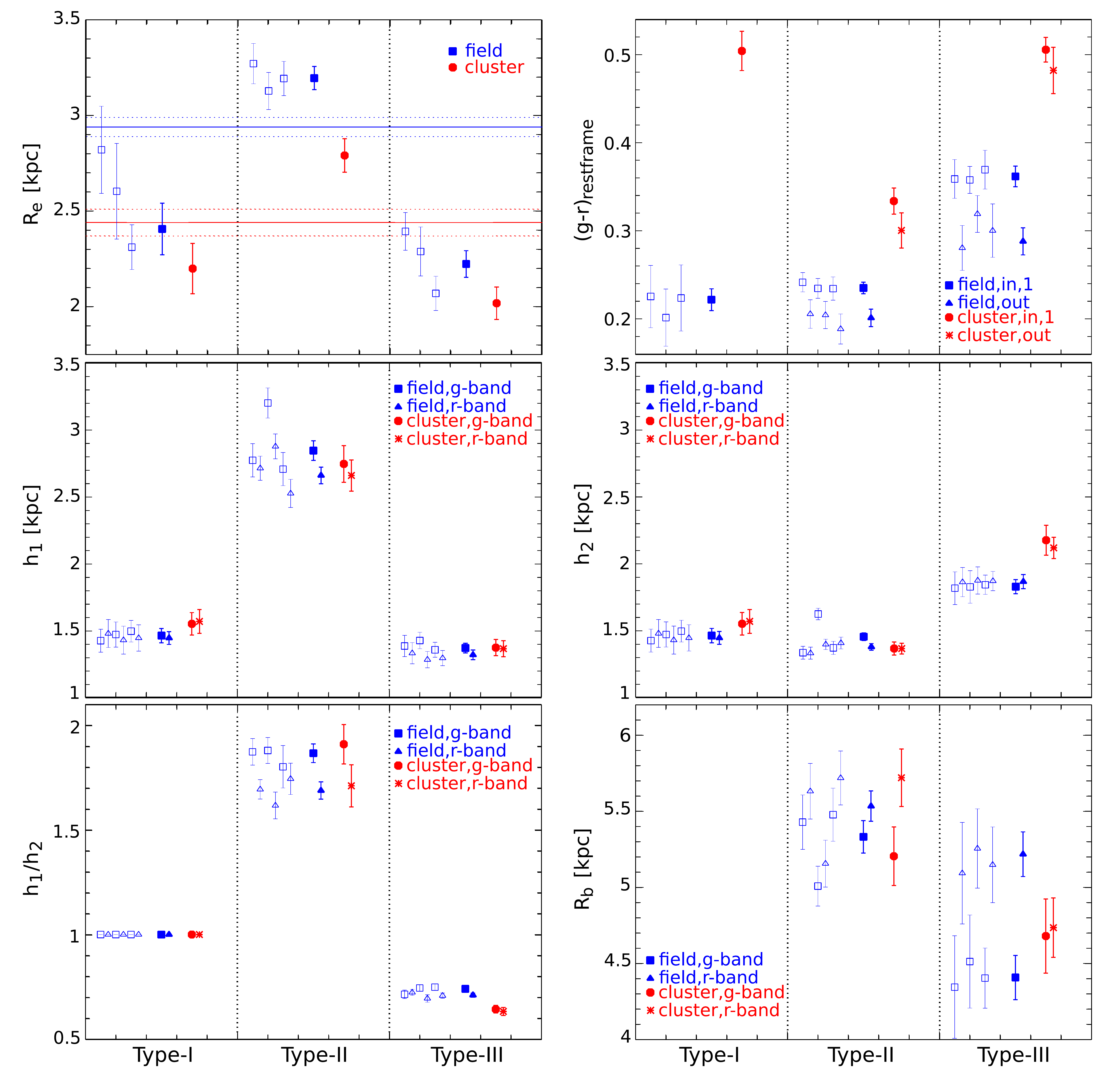}
\caption{Median galaxy properties for both environments and all three surface-brightness profile types. Errors are 1-sigma bootstrap estimates (1000 iterations). Open symbols and thin error bars represent the individual field samples (1, 2 and 3 from left to right, respectively). \textit{Top row:} effective radius, $R_{e}$, and $(g-r)_{restframe}$ colour. For Type II and Type III galaxies, $(g-r)_{in,1}$ and $(g-r)_{out}$ are plotted. For Type I galaxies, we plot only $(g-r)_{in,1}$, which in this case is the colour measured in the entire disc region (bulge masked at 0.5 $R_{e}$). The horizontal lines in the first panel indicate the median $R_{e}$ (and bootstrap errors) for the total cluster and total field sample, respectively. \textit{Middle row:} inner ($h_{1}$) and outer ($h_{2}$) exponential scale length for both measured bands. \textit{Bottom row:} Scale length ratio ($h_{1}$/$h_{2}$) and break radius ($R_{b}$) for both measured bands. For Type I galaxies, the break radius is not defined and the scale length ratio is by definition equal to unity.}
\label{fig:3}
\end{figure*}

There are several global results to highlight in this section. In relation to the global (break) structure: a) the number of Type I (single-exponential) galaxies in the cluster environment
is significantly higher (by a factor of 2.5) than in the field; b) the number of Type II (truncated) galaxies is lower in the cluster environment by 10 percentage points, matching the increase of Type I objects; c) the number of Type III (anti-truncated) galaxies, however, is quite similar in both environments. It is worth noting that the average radial distribution of the different surface-brightness profile types within the clusters is
different, with Type III galaxies residing at a median clustercentric distance (normalised to $r_{200}$) of 0.58$\pm$0.05, while the other types preferentially occupy regions at comparatively larger median normalised distances of 0.98$\pm$0.09 and 0.86$\pm$0.04 for Type I and II, respectively (errors estimated via 1000 1-$\sigma$ bootstrapping iterations). The normalised radial distribution of the different profile types is illustrated in Fig. \ref{fig:dist_norm}.
The global size value, as mentioned before, measured by $R_{e}$, is $\sim$15\% larger in the field than in the cluster. Finally, also the global
S\'{e}rsic-index (as provided by the NYU catalogue) is significantly higher in the cluster than in the field (by $\sim$15\%). By construction of our
subsamples, there is no global difference between cluster and field environment in average total stellar mass. The global structural and colour differences between the cluster and the field environment are illustrated in Fig. \ref{fig:firsthist}.\\  
\noindent Independent of the surface-brightness profile type, in all the regions, the median $(g-r)_{restframe}$ colour is significantly ($\sim$0.2
mag) redder in the cluster than in the field (most significant for Type I and III). For Type II and III field galaxies, with the bulge suitably masked, we find the inner disc to be notably redder than the regions at galactocentric distances greater than the break radius. In the cluster sample no such difference is seen within the errors. This finding indicates that the reddening in the cluster is stronger in the outer parts of the disc, which is in compliance with the scenario of disc-fading (see e.g. \citealt{christlein04}), i.e. the fading/aging of a galaxy's (outer) disc consequent to the stripping of gas by ram-pressure or to the consumption of gas through star formation (i.e. ``strangulation"). 
Since we have selected our samples to minimise the influence of dust (see Section \ref{sec:data}), we can assume that the observed reddening is truly caused by older stellar populations. Even though our data is not good enough to robustly determine the morphological type of the galaxies, we can further assume that the reddening is not solely related to a different morphological mix (i.e. a higher fraction of S0s) in the cluster environment; it has been shown that there exist hardly any Type II S0 galaxies in clusters (\citealt{erwin12,maltby15}), however, the reddening we detect is similarly seen in Type II and Type III galaxies. Moreover, given our selection limit in S\'{e}rsic-index and stellar mass, our sample is very likely biased towards late-type galaxies (see e.g. \citealt{ravindranath04}) and against S0s, which are among the most massive disc objects. The reddening observed in Type II and III cluster galaxies is slightly stronger in the outer disc regions, which is in compliance with galaxy evolutionary scenarios invoking ram-pressure stripping (see e.g. \citealt{steinhauser16}) or stellar migration (see e.g. \citealt{roskar08}). A summary of the global properties  discussed here is illustrated in Fig. \ref{fig:3}.

\begin{figure*}
\centering
\vspace{0.5cm}
\includegraphics[width=0.725\textwidth]{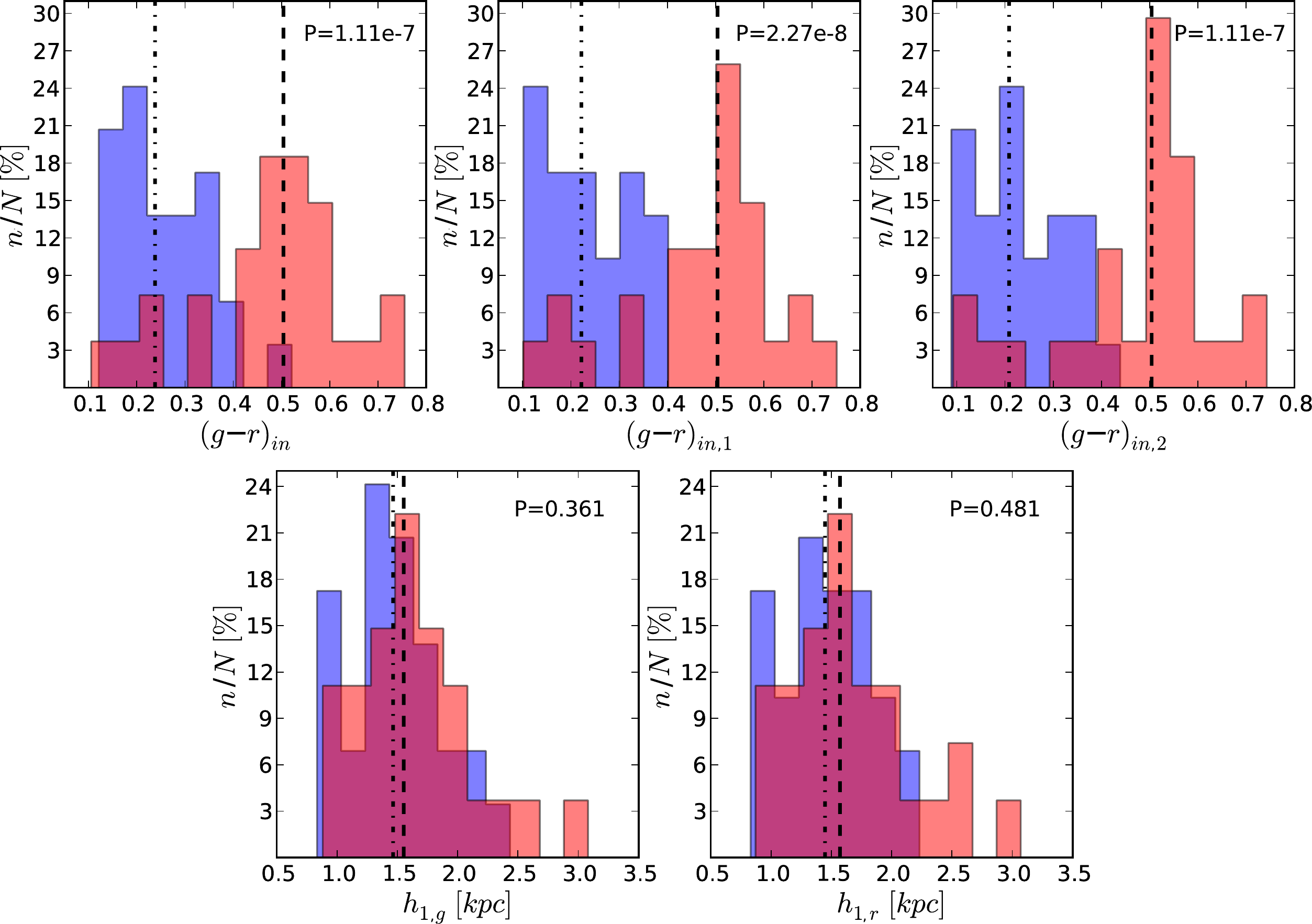}
\caption{Normalised distributions of the measured disc properties for Type I galaxies (field galaxies in blue, cluster galaxies in red). The vertical lines indicate the median values (dash-dotted for field and dashed for cluster galaxies). The KS-test P-value is shown in the upper right corner of each panel.}
\label{fig:hist1}
\end{figure*}

\begin{figure*}
\centering
\vspace{0.5cm}
\includegraphics[width=\textwidth]{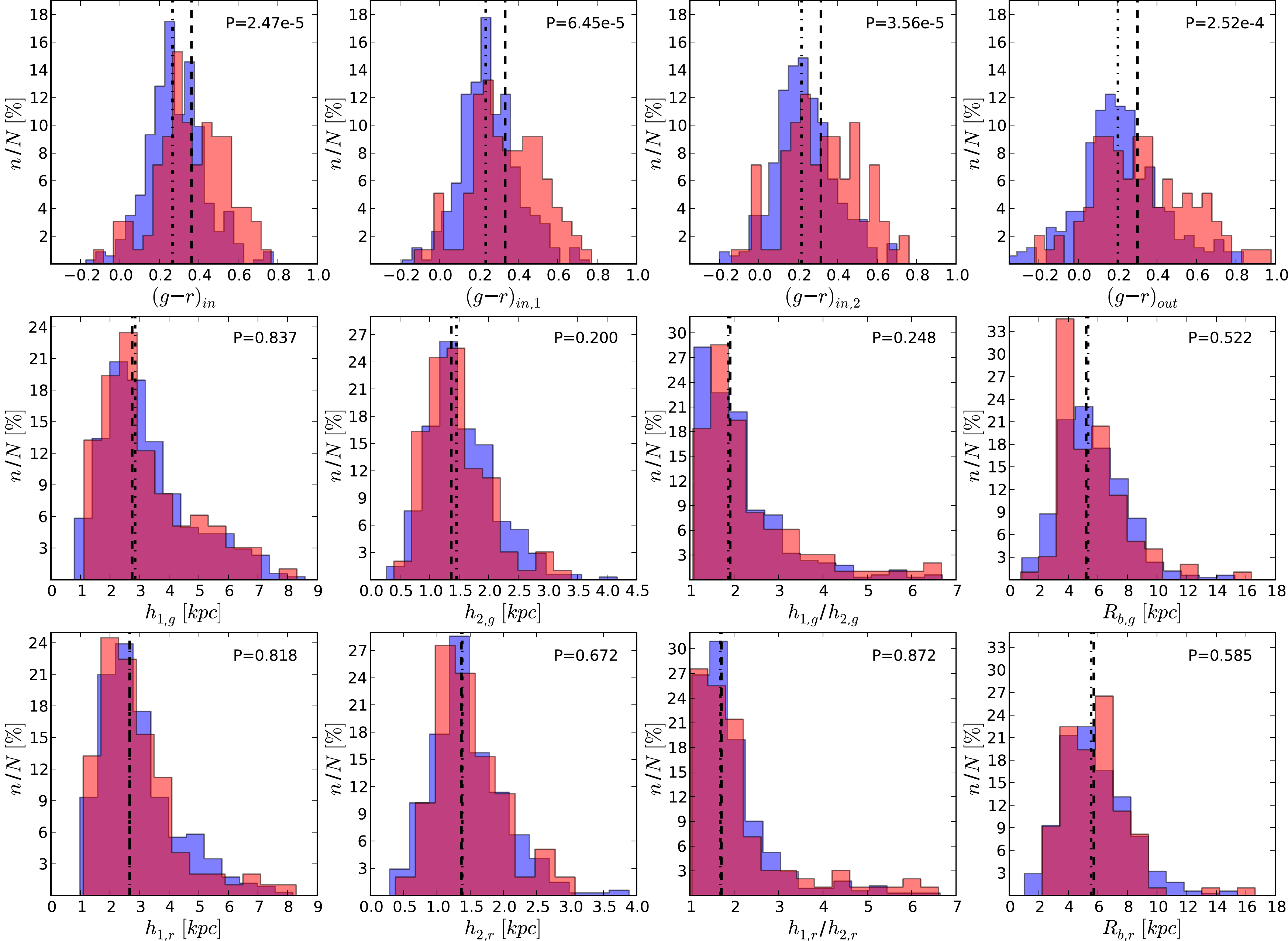}
\caption{Same as Fig. \ref{fig:hist1} for Type II galaxies.}
\label{fig:hist2}
\end{figure*}

\begin{figure*}
\centering
\includegraphics[width=\textwidth]{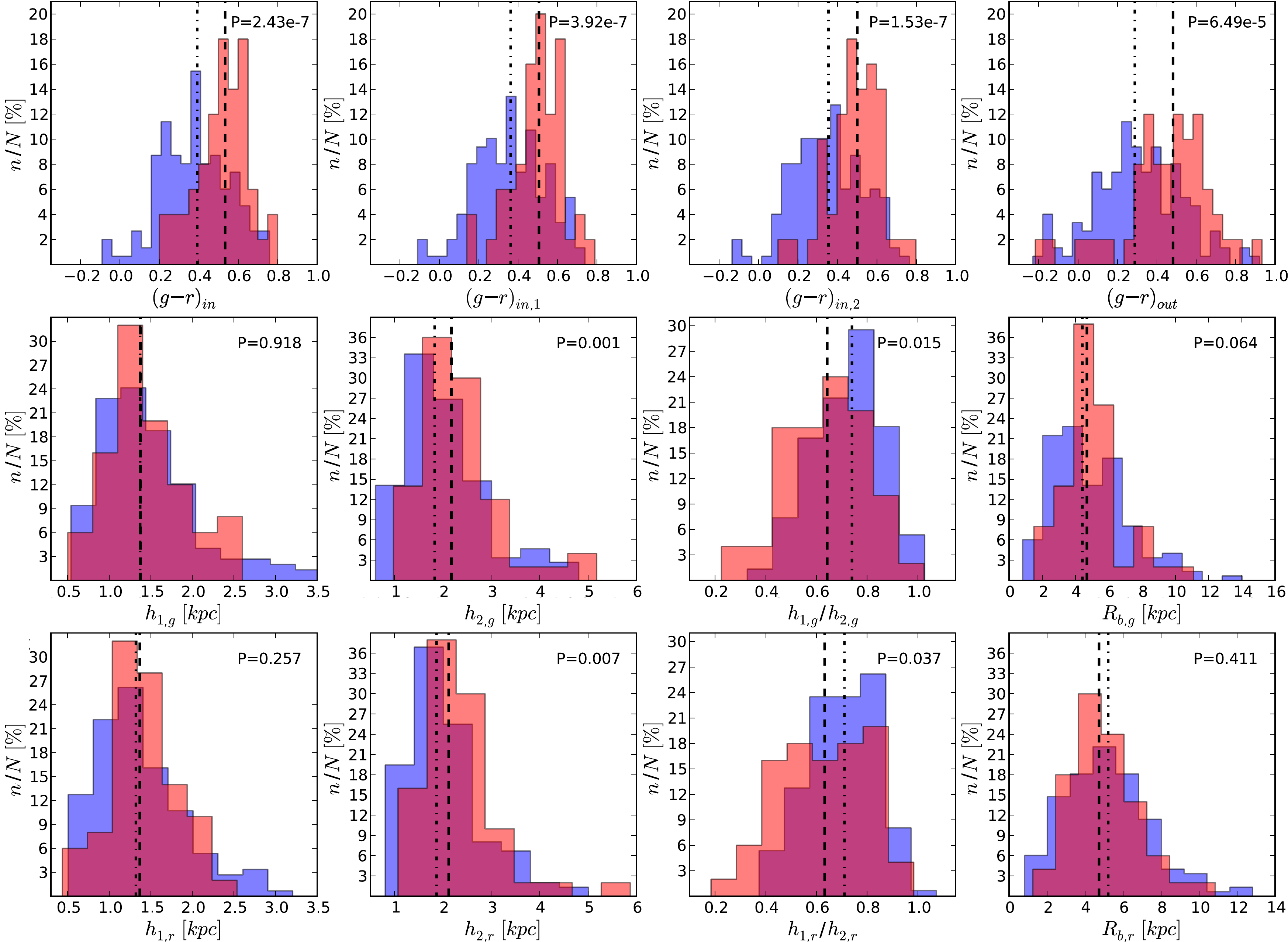}
\caption{Same as Fig. \ref{fig:hist1} for Type III galaxies.}
\label{fig:hist3}
\end{figure*}

\subsection{The different profile types}

In total, the fractions of Type I, II and III galaxies in the total cluster and field sample are 15$_{-4}^{+7}$\%, 56$_{-7}^{+7}$\% and 29$_{-7}^{+7}$\% and 6$_{-2}^{+2}$\%, 66$_{-4}^{+4}$\% and 29$_{-4}^{+4}$\%, respectively (errors correspond to \citet{wilson27}-confidence intervals). In Table \ref{table:results} we list these fractions along with the median values of the structural parameters for the galaxies in both environments (numbers with an asterisk indicate significant differences between cluster and field, according to a standard KS-test P-value $<$0.05). We plot the distributions of the measured parameters for the different profile types in Figs. \ref{fig:hist1} to \ref{fig:hist3} and illustrate their median values in Fig. \ref{fig:3}. The results for the three individual field samples can be found in Table \ref{table:resultsfc} for comparison. Note that all trends reported in this paper are confirmed by the evaluation of the individual field samples. 

\subsubsection{Type I galaxies}

The most interesting result in relation to this type of galaxies is the potentially larger scale length (a factor of $\sim$1.08) in the clusters environment compared to the field . However, this result is not significant according to standard KS-testing (P-value: 0.481 and 0.361 for r- and g-band, respectively). Moreover, Type I cluster galaxies are as red as Type III cluster galaxies while in the field Type I colours are comparable to Type II field objects (see Figs. \ref{fig:hist1} and \ref{fig:3}). Note that for Type I galaxies we measure the colour in the entire disc region. 

\subsubsection{Type II galaxies}

The inner and outer scale length, $h_1$ and $h_2$, (in both bands) are similar in clusters and in the field. In relation to the position of the break, $R_b$, there is not an obvious trend
depending on environment. The most remarkable issue is that the location of the break position is independent of the band for the field galaxies (within the errors) but rather different in the clusters. However, the observed difference is not significant according to standard KS-testing. The decrease in $R_e$ (field to cluster) is strongest for Type II cluster galaxies. For an illustration of these results, (see Figs. \ref{fig:hist2} and \ref{fig:3}).

\subsubsection{Type III galaxies}

Following the trend found in Type I galaxies, the effect of the cluster environment in Type III galaxies is to significantly enlarge
the value of the outer scale length, $h_2$ (KS-test P-values 0.007 and 0.001 for r- and g-band, respectively). For Type III galaxies, the increase is by a factor of $\sim$1.16. We also find that the ratio of the inner-to-outer scale lengths is smaller in the cluster environment by $\sim$15\% (for both bands). These findings could be interpreted as tidal effects consequent to galaxy harassment (see \citealt{moore96} for a detailed definition), as an increase of the rate of minor mergers building up the outer disc or as an increased contribution of an extended bulge component at larger galactocentric radii. However, the latter scenario is unlikely since it has been found to occur preferentially in S0 galaxies (see \citealt{maltby12b}) and our sample is biased against those. 
While merger events could also explain why Type III galaxies in the cluster are more massive than those in the field (see \ref{table:results}), it has to be noted that the merger rate is assumed to be very small in inner cluster regions (where we find Type III galaxies to reside; see Fig. \ref{fig:dist_norm}) due to high relative velocities. However, interactions of Type III galaxies with the intra-cluster medium and/or with other cluster galaxies (and the gravitational potential of the cluster) could have taken place in differently dense cluster regions over their comparatively long infall-time and thus might have been efficiently changing the galaxies' properties. The position of the break does not seem to be connected with the observed band for Type III cluster galaxies. This is different in the field, where $R_{b}$ is smaller in the g-band. 
However, the data on the break radii show a large scatter (resulting in large error bars), the observed differences in $R_{b}$ between the environments are not significant according to KS-testing and hence the results on $R_{b}$ have to be interpreted with caution.
Illustrations of the results on Type III galaxies are given in (Figs. \ref{fig:hist3} and \ref{fig:3}).\\  
The comparison of the three individual field samples shows that for some measured properties there are notable differences (e.g. in $R_{e}$ for Type I; in $h_{1,g}$, $h_{2,g}$ and $R_{b}$ for Type II). However, the KS-test results (comparing each of the field samples to the corresponding cluster sample) are consistent for all of these quantities, except for one sole outlier, namely $h_{2,g}$ for Type II galaxies in field sample 2.

\section{Discussion}
\label{sec:discussion}

Comparing two different environments like the field (low to intermediate density) and the cluster (high density), we are in position to explore how the physical
processes associated with high overdensity regions affect the peripheral parts of galaxy discs. There are three significant
differences between our field and cluster samples:

\begin{itemize}

\item The global size of the galaxies, as parameterised by the effective radius, is smaller by $\sim$15\% in the cluster
environment than in the field, while the global S\'ersic-index is higher in the cluster by $\sim$15\%.

\item The global (g-r) colour of the galaxies is redder by around 0.2 magnitudes in the clusters than in the field.

\item We find $\sim$2.5 times more Type I (pure exponential disc) galaxies in the clusters than in the field (15\% vs. 5\%). This difference is compensated by the lack of the corresponding percentage of Type II cluster galaxies (56\% vs. 66\%). 

\end{itemize}

\noindent Our work allows us to probe, in detail, how and where the ageing and global size transformations have taken place. Figure
\ref{fig:3} shows that the global size difference between the discs in the field and in the clusters, even though largest in Type II galaxies, seems to hold for all the profile types when explored separately. When taking a detailed look
at the reddening of the individual profile types, Fig. \ref{fig:3} indicates that all the classes have undergone a
similar reddening from the field to the cluster environments. It is worth noting that both Type I and II have a similar colour in the field, the Type III objects being notably redder. However, in the cluster environment this colour
similarity is reversed so that Type I and III are the ones which share similar colours. It is remarkable that Type III field galaxies are redder than the other profile types even after masking the bulge. A possible explanation is that Type III field galaxies are formed by tidal interactions and minor mergers in the course of group pre-processing (see e.g. \citealt{younger07, lopes14}) which might have led to the observed reddening. The fact that Type II cluster galaxies are bluer than the other types in the same environment could indicate that the Type II feature is erased relatively quickly upon the cluster infall, allowing only for a limited reddening before the break vanishes. Note, however, that these are tentative interpretations.\\
\noindent As the inner scale length of the disc galaxies, $h_1$, barely changes when moving from the field to the clusters, to
understand the ultimate reason of the change in the global size of the galaxies we need to focus our attention on the outer
scale length, $h_2$. For Type I and Type III discs, we find that $h_2$ is larger (by a factor of $\sim$1.08 and $\sim$1.16, respectively) in
the clusters than in the field. If this was the only difference in the structure of the field versus the cluster
galaxies, the global size of the cluster galaxies (at fixed stellar mass) should be larger than in the field. To
understand why the cluster galaxies are yet more compact, we have to account also for the reddening of the objects. At
decreasing the brightness of the galaxies' discs, the bulges of all these objects become more prominent. This effect
moves the effective radius towards the inner regions of the objects.\\
\noindent Both the increment of $h_2$ and the global reddening of the discs in the cluster regions are suggestive of physical
process connected to the cluster environment. 
The rise of $h_2$ in Type I and more significantly in Type III cluster galaxies could be understood as the result of an increased contribution of a prominent bulge component. Since we already accounted for the bulge component when fitting the surface-brightness profiles, this hypothesis is not sufficient to explain our findings. Other processes that might play an important role include tidal effects consequent to galaxy harassment (\citealt{moore96}) and minor merger events building up the outer disc. The global reddening of the cluster galaxies is potentially connected to the decrease
of star formation in the cluster associated with the exhaustion of gas by ram-pressure stripping (see e.g.
\citealt{boesch13, pranger13, steinhauser16}). It is worth noting that the reddening of the discs seems to be stronger outside the break radius of the galaxies. This is consistent with the view that gas removal by ram-pressure stripping should be more efficient in the outer disc regions (see e.g. \citealt{steinhauser12}).\\
\indent Another interesting aspect to discuss is the dramatic increase (by a factor of $\sim$2.5) in the number of Type I discs in
the clusters. This large variation in the frequency of Type I galaxies is accompanied by a substantial change in their
colours. Whereas Type II and III have become redder by around 0.1-0.2 magnitudes, in the case of Type I this change is
$\sim$0.3 mag. The large increase in the number of Type I galaxies is compensated by a corresponding decrease of the percentage of Type II objects. This, together with our results on galaxy colour, indicates a transformation from Type II to Type I as the galaxies fall onto the cluster.
This hypothesis is in compliance with \citet{maltby15} who find indications for a transformation from spiral to S0 as the break is erased in Type II galaxies that fall onto a cluster, however, \citet{maltby15} (and \citet{maltby12b}) conclude that the structural disc parameters are not significantly influenced by environment. This disparity could be explained by the higher mass and redshift range of the galaxies selected for their study that might have made it more difficult to detect environmental trends. Furthermore, our suggested scenario is in agreement with \citet{erwin12} who analysed S0 galaxies in the cluster and in the field and concluded that the lack of Type II cluster S0s indicates a transformation from Type II to Type I. A suppression of the survival of Type II objects in the cluster environment has also been reported by \citet{roediger12}, analysing Virgo disc galaxies and, recently, by \citet{clarke17} using N-body SPH simulations. Supporting this view, it is worth noting that the scale length of the Type I discs has a value which is
intermediate between the corresponding Type II and Type III values (both for the inner and outer scalengths). This has also been found by
\citealt{munoz-mateos13}. Also their comparatively large scatter in $R_{e}$ can be interpreted as a hint for Type I galaxies experiencing a structural transformation. The unchanging frequency of Type III galaxies from field to cluster remains unexplained by our analysis and interpretation. We will address this point in our follow-up investigations.\\
Our results also have to be interpreted with respect to galaxy morphology. The quality of our data is not high enough to robustly assign all objects to morphological classes along the Hubble sequence, however, invoking the morphology-density relation (\citealt{dressler80}) we can assume a higher fraction of early-type disc galaxies in the cluster environment. Referring to \citet{gutierrez11} who find that Type I and III galaxies are more common in early-type discs while Type II galaxies are dominant in late-type discs, this would indicate that the trends we illustrate in Fig. \ref{fig:3} could - at least to some extent - be caused by morphological segregation between the two environments. On the other hand, as explained in Section \ref{sec:results}, our sample is very likely biased against S0 galaxies due to our selection criteria in stellar mass and S\'ersic-index. Moreover, the reddening for cluster Type II and III profiles is approximately the same. Since Type II galaxies are not found in cluster S0s (\citealt{erwin12, maltby15}), this indicates that the trends we find are not driven by S0 galaxies. In general, we conclude that our results are not significantly influenced by morphological segregation.

\section{Summary}
\label{sec:summary}
We have selected four samples of disc galaxies within the same narrow stellar mass range (0.8$<$ $M_{\star}$ $<$ 4) $\times$
10$^{10}$ $M_{\odot}$. While one of these samples consists of galaxies residing in galaxy clusters, the other three consist of galaxies outside of galaxy clusters, i.e. from the field. Each of the four samples holds around 175 galaxies which have been classified according to their disc
structure (Type I
$\equiv$ single-exponential, Type II $\equiv$ truncated, Type III $\equiv$ anti-truncated). We
find the following results:

\begin{itemize}
\item Disc galaxies are $\sim$15\% more compact in clusters than in the
field. They are also $\sim$0.2 mag redder in (g-r) colour and show higher S\'ersic-indices (by $\sim$15\%) in the cluster region. The reddening is observed both inside and outside the break radius.\\
\item We find $\sim$2.5 times more Type I (pure exponential disc) galaxies in the clusters than in the field. This difference is compensated by the lack of the corresponding percentage of Type II cluster galaxies. The fraction of Type III galaxies is the same in both environments.\\
\item Type III cluster galaxies reside significantly closer to the cluster centre than the other break types.\\
\item The structural parameter that changes
most significantly, at comparing cluster versus field, is the outer scalelength of Type III discs, increasing by $\sim$16\% from field to cluster. Consequent to this change, the ratio of the inner and outer scale length of Type III galaxies changes by $\sim$15\%.\\
\item We suggest that Type I galaxies form from Type II galaxies, consequent to the physical mechanisms acting on the Type II population, produced/enhanced by environment. \\
\end{itemize} 
  
\noindent In a follow-up to this work we will investigate the structural parameters of disc galaxies as functions of clustercentric distance. To ensure adequate statistics, our analyses will be carried out on an extended sample of galaxies, including a representative fraction of galaxies that reside in the transition region between cluster and field. For a consistent follow-up we will apply the same methods as presented and described in this paper.\\

\noindent A selection of prototypical surface-brightness profiles, the evaluation of the field control samples and comprehensive lists of the measured and analysed galaxy data can be found in the appendix. 

\section*{Acknowledgements}
We would like to thank the anonymous referee for detailed comments that contributed a lot to improving the quality of the present article. IT and MC acknowledge support from grant AYA2013-48226-C3-1-P and programme SEV-2011-0187 from the Spanish Ministry of Economy and Competitiveness (MINECO). This work is part of the research activities at the National Astronomical Research Institute of Thailand (Public Organization).

\bibliographystyle{mnras}
\bibliography{bib_cluster_disks}
\bsp
\onecolumn
\newpage
\appendix
\captionsetup{labelfont=bf}
\section{Prototypical profiles}

\begin{center}
\includegraphics[width=0.88\textwidth]{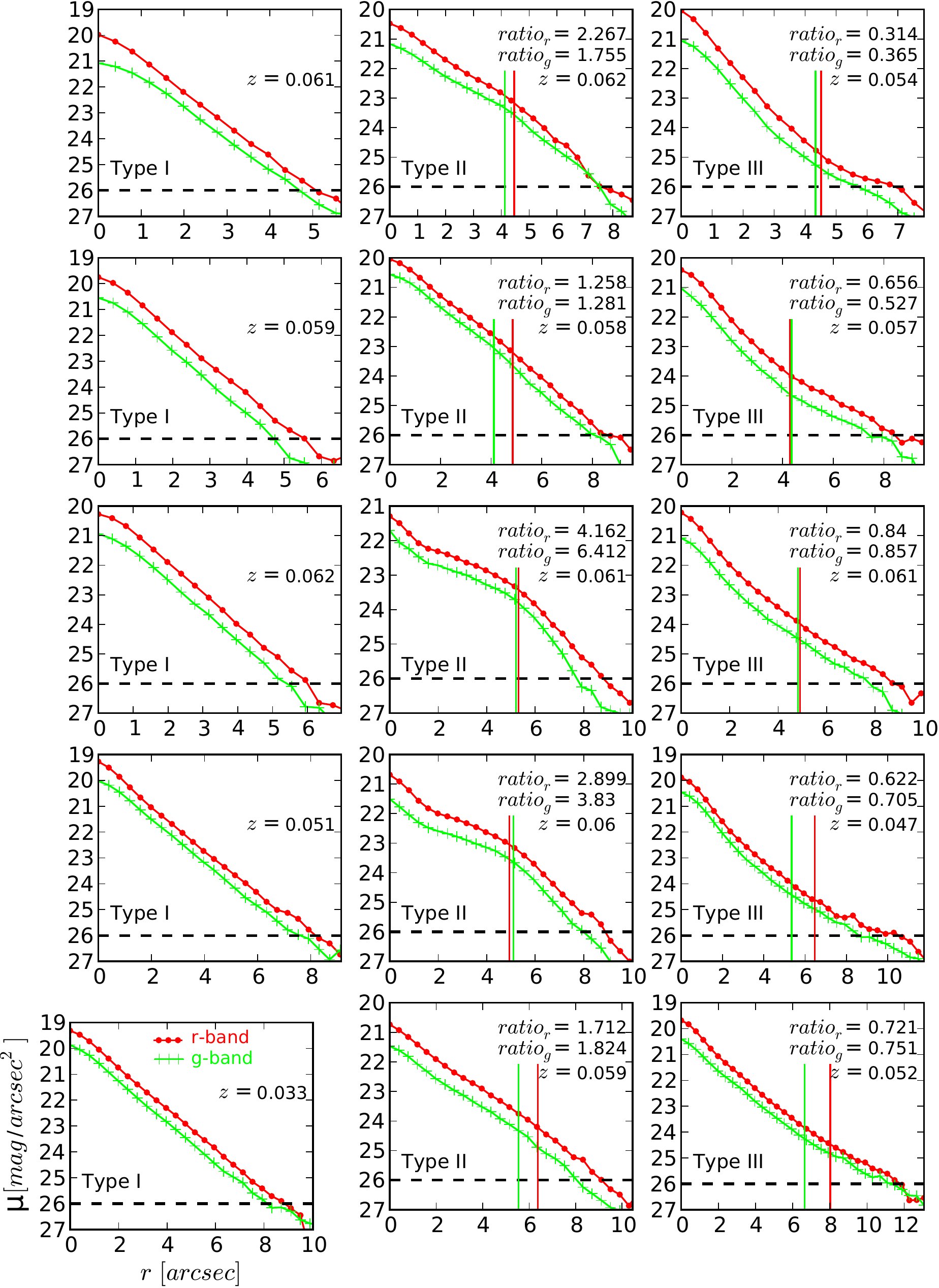}
\captionof{figure}{Prototypical cluster galaxy profiles as obtained and analysed in this work, sorted from top to bottom according to apparent size. The left, middle and right column contain Type I, Type II and Type III profiles, respectively. Vertical short lines indicate the break radii in the g-band (green) and r-band (red), the horizontal dashed line marks our conservative confidence threshold. In addition, galaxy redshifts and measured scale length ratios ($h_{1}$/$h_{2}$) are given.}
\label{fig:A1}
\end{center}

\newpage
\begin{center}

\includegraphics[width=0.88\textwidth]{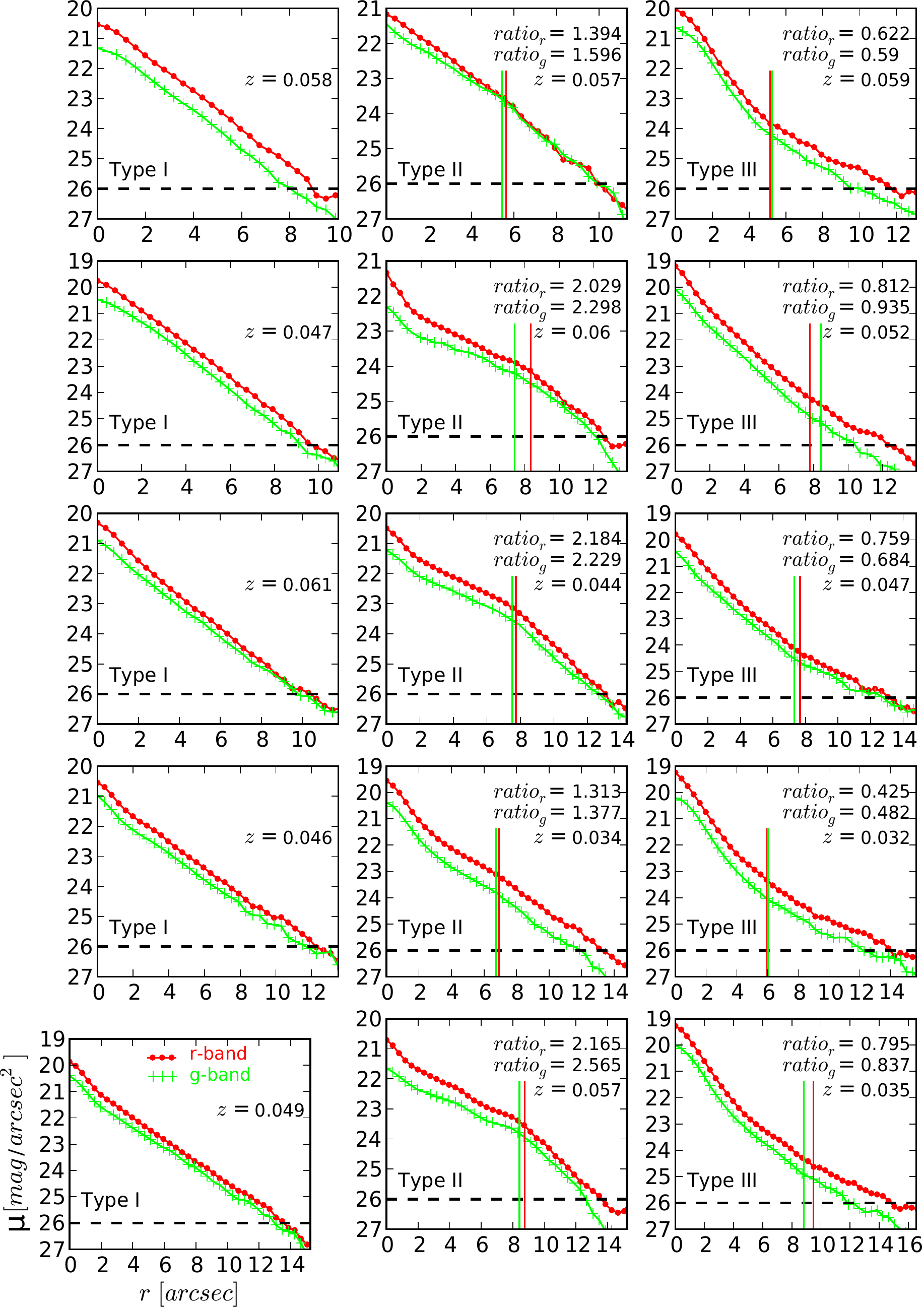}
\captionof{figure}{Same as Fig. \ref{fig:A1} for field galaxies.}
\label{fig:A2}
\end{center}

\newpage
\section{Analysis of the separated field samples}
\captionof{table}{Listing of the median values of the structural properties of the galaxies in the three initially selected field samples. See Table \ref{table:results} for a detailed description.}
\label{table:resultsfc}

\begin{center}
\begin{tabular}{c|ccc|c}
\hline
\hline
Sample & Type I & Type II & Type III & all types \\
\hline
\multicolumn{1}{c|}{} &
\multicolumn{4}{c}{Field 1} \\
\hline
N & 10 & 115 & 47 & 172 \\
\% & 6$_{-3}^{+4}$ & 67$_{-8}^{+6}$ & 27$_{-6}^{+7}$ & 100 \\
\hline
$R_{e}$ [kpc] & 2.82$\pm$0.23 & 3.27$\pm$0.11* & 2.39$\pm$0.10 & 3.10$\pm$0.09* \\
$n$ & 1.81$\pm$0.07 & 1.60$\pm$0.03 & 2.05$\pm$0.06* & 1.74$\pm$0.03* \\
$M_{\star}$ [10$^{10}M_{\odot}$] & 1.69$\pm$0.16 & 1.62$\pm$0.06 & 1.55$\pm$0.08 & 1.60$\pm$0.05 \\
\hline
$(g-r)_{in}$ & 0.235$\pm$0.041* & 0.282$\pm$0.010* & 0.379$\pm$0.022* & 0.283$\pm$0.010* \\
$(g-r)_{in,1}$ & 0.225$\pm$0.035* & 0.242$\pm$0.011* & 0.359$\pm$0.022* & 0.270$\pm$0.010* \\
$(g-r)_{in,2}$ & 0.214$\pm$0.037* & 0.223$\pm$0.010* & 0.355$\pm$0.022* & 0.249$\pm$0.010* \\
$(g-r)_{out}$ & - & 0.199$\pm$0.016* & 0.281$\pm$0.025* & 0.232$\pm$0.015* \\
\hline
$h_{1,g}$ [kpc] & 1.43$\pm$0.09 & 2.77$\pm$0.12 & 1.39$\pm$0.08 & - \\
$h_{2,g}$ [kpc] & -"- & 1.34$\pm$0.05 & 1.82$\pm$0.12* & - \\
$ratio_{g}$ & 1 & 1.87$\pm$0.06 & 0.71$\pm$0.02 & - \\
$R_{b,g}$ [kpc] & - & 5.43$\pm$0.18 & 4.35$\pm$0.33 & - \\
$h_{1,r}$ [kpc] & 1.48$\pm$0.10 & 2.72$\pm$0.09 & 1.33$\pm$0.08 & - \\
$h_{2,r}$ [kpc] & -"- & 1.33$\pm$0.05 & 1.86$\pm$0.11 & - \\
$ratio_{r}$ & 1 & 1.69$\pm$0.05 & 0.72$\pm$0.01* & - \\
$R_{b,r}$ [kpc] & - & 5.63$\pm$0.18 & 5.09$\pm$0.33 & - \\
\hline
\hline
\multicolumn{1}{c|}{} &
\multicolumn{4}{c}{Field 2} \\
\hline
N & 10 & 112 & 50 & 172 \\
\% & 6$_{-3}^{+4}$ & 65$_{-7}^{+7}$ & 29$_{-6}^{+7}$ & 100 \\
\hline
$R_{e}$ [kpc] & 2.60$\pm$0.25 & 3.13$\pm$0.10 & 2.29$\pm$0.13 & 2.91$\pm$0.08* \\
$n$ & 1.60$\pm$0.11 & 1.53$\pm$0.03* & 1.94$\pm$0.04* & 1.68$\pm$0.03* \\
$M_{\star}$ [10$^{10}M_{\odot}$] & 1.47$\pm$0.18 & 1.60$\pm$0.05 & 1.68$\pm$0.09 & 1.61$\pm$0.05 \\
\hline
$(g-r)_{in}$ & 0.214$\pm$0.031* & 0.256$\pm$0.011* & 0.389$\pm$0.016* & 0.275$\pm$0.010* \\
$(g-r)_{in,1}$ & 0.201$\pm$0.032* & 0.235$\pm$0.011* & 0.358$\pm$0.015* & 0.253$\pm$0.011* \\
$(g-r)_{in,2}$ & 0.202$\pm$0.033* & 0.213$\pm$0.012* & 0.348$\pm$0.015* & 0.242$\pm$0.010* \\
$(g-r)_{out}$ & - & 0.197$\pm$0.015* & 0.302$\pm$0.021* & 0.215$\pm$0.013* \\
\hline
$h_{1,g}$ [kpc] & 1.47$\pm$0.09 & 3.20$\pm$0.11 & 1.43$\pm$0.06 & - \\
$h_{2,g}$ [kpc] & -"- & 1.42$\pm$0.04* & 1.83$\pm$0.12* & - \\
$ratio_{g}$ & 1 & 1.88$\pm$0.06 & 0.74$\pm$0.02* & - \\
$R_{b,g}$ [kpc] & - & 5.01$\pm$0.15 & 4.51$\pm$0.26 & - \\
$h_{1,r}$ [kpc] & 1.43$\pm$0.10 & 2.88$\pm$0.09 & 1.29$\pm$0.06 & - \\
$h_{2,r}$ [kpc] & -"- & 1.40$\pm$0.04 & 1.88$\pm$0.10* & - \\
$ratio_{r}$ & 1 & 1.62$\pm$0.06 & 0.69$\pm$0.02 & - \\
$R_{b,r}$ [kpc] & - & 5.16$\pm$0.15 & 5.26$\pm$0.26 & - \\
\hline
\hline
\multicolumn{1}{c|}{} &
\multicolumn{4}{c}{Field 3}\\
\hline
N & 9 & 116 & 52 & 177 \\
\% & 5$_{-2}^{+4}$ & 66$_{-8}^{+6}$ & 29$_{-6}^{+7}$ & 100 \\
\hline
$R_{e}$ [kpc] & 2.31$\pm$0.12 & 3.19$\pm$0.09 & 2.07$\pm$0.09 & 2.74$\pm$0.07* \\
$n$ & 1.33$\pm$0.09* & 1.71$\pm$0.03 & 2.01$\pm$0.05* & 1.76$\pm$0.03* \\
$M_{\star}$ [10$^{10}M_{\odot}$] & 1.55$\pm$0.12 & 1.59$\pm$0.05 & 1.50$\pm$0.08 & 1.56$\pm$0.05 \\
\hline
$(g-r)_{in}$ & 0.238$\pm$0.038* & 0.269$\pm$0.012* & 0.401$\pm$0.021* & 0.294$\pm$0.012* \\
$(g-r)_{in,1}$ & 0.224$\pm$0.038* & 0.234$\pm$0.013* & 0.369$\pm$0.022* & 0.271$\pm$0.012* \\
$(g-r)_{in,2}$ & 0.224$\pm$0.037* & 0.213$\pm$0.014* & 0.351$\pm$0.020* & 0.252$\pm$0.013* \\
$(g-r)_{out}$ & - & 0.189$\pm$0.017* & 0.300$\pm$0.030* & 0.238$\pm$0.015* \\
\hline
$h_{1,g}$ [kpc] & 1.49$\pm$0.08 & 2.71$\pm$0.12 & 1.36$\pm$0.06 & - \\
$h_{2,g}$ [kpc] &  -"- & 1.37$\pm$0.05 & 1.84$\pm$0.07* & - \\
$ratio_{g}$ & 1 & 1.80$\pm$0.10 & 0.75$\pm$0.01* & - \\
$R_{b,g}$ [kpc] & - & 5.48$\pm$0.18 & 4.40$\pm$0.25 & - \\
$h_{1,r}$ [kpc] & 1.45$\pm$0.10 & 2.53$\pm$0.10 & 1.30$\pm$0.06 & - \\
$h_{2,r}$ [kpc] & -"- & 1.41$\pm$0.04 & 1.87$\pm$0.07* & - \\
$ratio_{r}$ & 1 & 1.74$\pm$0.07 & 0.71$\pm$0.01 & - \\
$R_{b,r}$ [kpc] & - & 5.72$\pm$0.18 & 5.15$\pm$0.25 & - \\
\hline
\hline
\end{tabular} 

\vspace{4.5cm}

\includegraphics[width=0.75\textwidth]{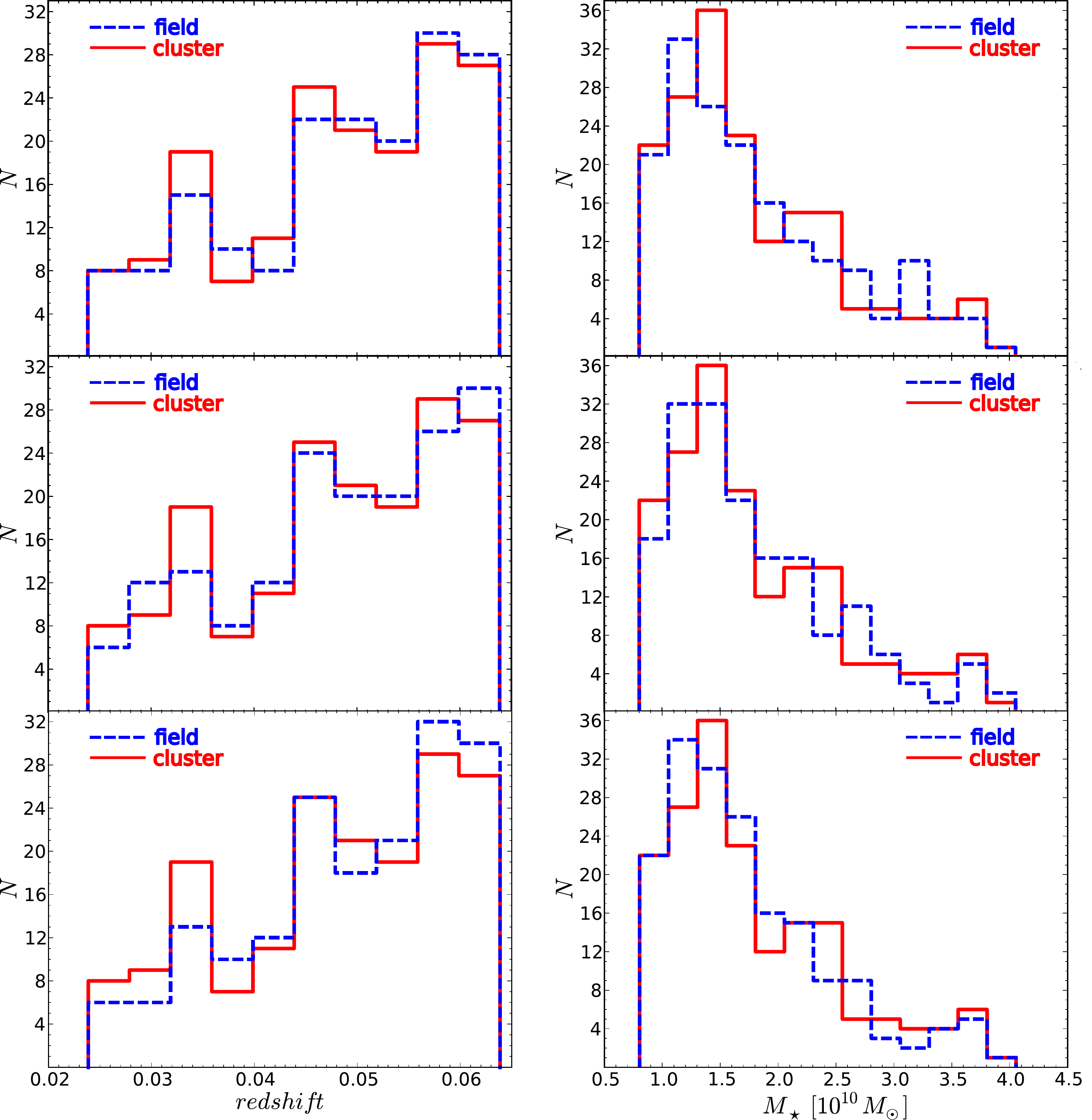}
\captionof{figure}{Redshift (left) and stellar mass (right) distributions of the total cluster sample and the three field samples (field sample 1, 2 and 3 from top to bottom).}
\label{fig:1a}




\end{center}
\newpage

\section{Galaxy data}
\label{sec:galdat}
\captionof{table}{Listing of coordinates, profile type, redshift, effective radius, S\'ersic-index, colours and NYU-VAGC-ID of all 175 galaxies in the total cluster sample.}
\centering
\begin{tabular}{cccccccccccc}
\hline
\hline
\multicolumn{12}{c}{Total cluster sample} \\
\hline
Index & RA. [deg] & DEC. [deg] & Type & z & $R_{e}$ [kpc] & $n$ & $(g-r)_{in}$ & $(g-r)_{in,1}$ & $(g-r)_{in,2}$ & $(g-r)_{out}$ & ID\\
\hline
1	&	223.306	&	16.9354	&	I	&	0.0453	&	1.646	&	2.344	&	0.4951	&	0.4910	&	0.4882	&	-	&	2277479	\\
2	&	207.954	&	37.5459	&	I	&	0.0615	&	2.798	&	1.990	&	0.4249	&	0.4139	&	0.4037	&	-	&	1837246	\\
3	&	227.763	&	7.6391	&	I	&	0.0457	&	3.225	&	2.130	&	0.4338	&	0.4338	&	0.4338	&	-	&	1340835	\\
4	&	239.843	&	17.9775	&	I	&	0.0465	&	3.245	&	1.768	&	0.3342	&	0.3342	&	0.3342	&	-	&	1998978	\\
5	&	137.110	&	52.1416	&	I	&	0.0613	&	1.247	&	2.075	&	0.5695	&	0.5681	&	0.5664	&	-	&	0374342	\\
6	&	235.227	&	21.7966	&	I	&	0.0416	&	1.846	&	2.445	&	0.5414	&	0.5414	&	0.5414	&	-	&	1965469	\\
\hline
\hline
\end{tabular}
\label{tab:a1}
\newpage
\centering

\begin{tabular}{cccccccccccc}
\hline
\hline
\multicolumn{12}{c}{Total cluster sample (continued)} \\
\hline
Index & RA. [deg] & DEC. [deg] & Type & z & $R_{e}$ [kpc] & $n$ & $(g-r)_{in}$ & $(g-r)_{in,1}$ & $(g-r)_{in,2}$ & $(g-r)_{out}$ & ID\\
\hline
60	&	167.004	&	44.1476	&	II	&	0.0589	&	3.830	&	1.918	&	0.5561	&	0.5370	&	0.5330	&	0.5913	&	1248630	\\
61	&	213.736	&	4.9446	&	II	&	0.0562	&	2.968	&	1.378	&	0.3315	&	0.3223	&	0.3043	&	0.3569	&	0306768	\\
62	&	136.784	&	52.1685	&	II	&	0.0623	&	5.713	&	1.824	&	0.1104	&	0.0661	&	0.0414	&	0.0444	&	0432310	\\
63	&	128.780	&	38.2247	&	II	&	0.0569	&	5.149	&	2.192	&	0.2922	&	0.2498	&	0.2342	&	0.3275	&	0804407	\\
64	&	167.120	&	44.2202	&	II	&	0.0588	&	3.199	&	1.580	&	0.4344	&	0.4269	&	0.4186	&	0.5332	&	1248639	\\
65	&	203.714	&	34.0555	&	II	&	0.0251	&	2.870	&	1.659	&	0.5307	&	0.5003	&	0.4947	&	0.3846	&	1928823	\\
\hline
\hline
\end{tabular}
\newpage

\begin{tabular}{cccccccccccc}
\hline
\hline
\multicolumn{12}{c}{Total cluster sample (continued)} \\
\hline
Index & RA. [deg] & DEC. [deg] & Type & z & $R_{e}$ [kpc] & $n$ & $(g-r)_{in}$ & $(g-r)_{in,1}$ & $(g-r)_{in,2}$ & $(g-r)_{out}$ & ID\\
\hline
119	&	205.241	&	29.8189	&	II	&	0.0433	&	2.983	&	1.383	&	0.0524	&	0.0193	&	-0.0075	&	-0.0126	&	1936820	\\
120	&	218.608	&	52.9602	&	II	&	0.0452	&	4.647	&	1.532	&	0.2610	&	0.2135	&	0.1893	&	0.1116	&	1191490	\\
121	&	167.395	&	21.6654	&	II	&	0.0317	&	1.967	&	1.531	&	0.3743	&	0.3490	&	0.3411	&	0.2330	&	2306478	\\
122	&	230.383	&	7.8144	&	II	&	0.0453	&	2.311	&	1.461	&	0.4713	&	0.4632	&	0.4636	&	0.5016	&	1343316	\\
123	&	203.894	&	36.7494	&	II	&	0.0615	&	4.136	&	1.072	&	0.2907	&	0.2486	&	0.2271	&	0.2091	&	1879979	\\
124	&	200.225	&	31.2495	&	II	&	0.0452	&	3.729	&	1.535	&	0.2155	&	0.1844	&	0.1559	&	0.1109	&	1948165	\\
\hline
\hline
\end{tabular}
\newpage
\captionof{table}{Listing of coordinates, profile type, redshift, effective radius, S\'ersic-index, colours and NYU-VAGC-IDof all 172 galaxies in field sample 1.}
\centering
\begin{tabular}{cccccccccccc}
\hline
\hline
\multicolumn{12}{c}{Field sample 1} \\
\hline
Index & RA. [deg] & DEC. [deg] & Type & z & $R_{e}$ [kpc] & $n$ & $(g-r)_{in}$ & $(g-r)_{in,1}$ & $(g-r)_{in,2}$ & $(g-r)_{out}$ & ID \\
\hline
1	&	135.394	&	2.3800	&	I	&	0.0609	&	1.274	&	1.598	&	0.2026	&	0.1990	&	0.1940	&	-	&	0452293	\\
2	&	165.776	&	5.2676	&	I	&	0.0517	&	3.028	&	2.193	&	0.4791	&	0.3620	&	0.3592	&	-	&	0450752	\\
3	&	210.160	&	18.8510	&	I	&	0.0613	&	2.614	&	1.785	&	0.1842	&	0.1663	&	0.1503	&	-	&	2365545	\\
4	&	148.309	&	22.5635	&	I	&	0.0620	&	3.949	&	1.823	&	0.1600	&	0.1643	&	0.1577	&	-	&	2235614	\\
5	&	244.952	&	34.3156	&	I	&	0.0333	&	1.345	&	1.578	&	0.3092	&	0.3056	&	0.3053	&	-	&	0984438	\\
6	&	146.060	&	24.1013	&	I	&	0.0515	&	1.728	&	1.999	&	0.2683	&	0.2518	&	0.2338	&	-	&	2204509	\\
\hline
\hline
\end{tabular}
\newpage
\centering

\begin{tabular}{cccccccccccc}
\hline
\hline
\multicolumn{12}{c}{Field sample 1 (continued)} \\
\hline
Index & RA. [deg] & DEC. [deg] & Type & z & $R_{e}$ [kpc] & $n$ & $(g-r)_{in}$ & $(g-r)_{in,1}$ & $(g-r)_{in,2}$ & $(g-r)_{out}$ & ID\\
\hline
60	&	243.436	&	51.0830	&	II	&	0.0474	&	3.111	&	2.110	&	0.4440	&	0.3918	&	0.3681	&	0.4065	&	1958893	\\
61	&	225.951	&	24.5416	&	II	&	0.0583	&	2.573	&	1.440	&	0.1958	&	0.1857	&	0.1826	&	0.2164	&	1954643	\\
62	&	166.244	&	31.4176	&	II	&	0.0468	&	3.960	&	1.705	&	0.1621	&	0.1404	&	0.1375	&	-0.0322	&	1962270	\\
63	&	165.740	&	31.3927	&	II	&	0.0461	&	4.507	&	1.893	&	0.2330	&	0.1799	&	0.1446	&	-0.3618	&	1962260	\\
64	&	143.999	&	26.6740	&	II	&	0.0336	&	1.887	&	2.161	&	0.2530	&	0.2345	&	0.2187	&	0.2141	&	2136922	\\
65	&	138.720	&	23.4598	&	II	&	0.0349	&	3.383	&	1.764	&	0.1229	&	0.0725	&	0.0454	&	-0.1470	&	2134713	\\
\hline
\hline
\end{tabular}
\newpage

\begin{tabular}{cccccccccccc}
\hline
\hline
\multicolumn{12}{c}{Field sample 1 (continued)} \\
\hline
Index & RA. [deg] & DEC. [deg] & Type & z & $R_{e}$ [kpc] & $n$ & $(g-r)_{in}$ & $(g-r)_{in,1}$ & $(g-r)_{in,2}$ & $(g-r)_{out}$ & ID\\
\hline
119	&	132.199	&	30.8083	&	II	&	0.0539	&	4.814	&	1.366	&	0.2011	&	0.1768	&	0.1667	&	0.1753	&	1087420	\\
120	&	158.216	&	43.6308	&	II	&	0.0537	&	9.824	&	2.050	&	0.1944	&	0.1499	&	0.1267	&	0.2402	&	1064537	\\
121	&	185.665	&	55.0830	&	II	&	0.0367	&	3.312	&	2.125	&	0.0917	&	0.0687	&	0.0555	&	0.0625	&	0964454	\\
122	&	184.269	&	55.1235	&	II	&	0.0521	&	3.037	&	1.192	&	0.3138	&	0.2909	&	0.2800	&	0.3792	&	0964422	\\
123	&	200.705	&	49.8553	&	II	&	0.0595	&	3.484	&	1.213	&	0.1100	&	0.0736	&	0.0473	&	-0.2302	&	1162715	\\
124	&	172.780	&	56.5250	&	II	&	0.0493	&	6.029	&	1.957	&	0.4247	&	0.3096	&	0.2738	&	-0.2185	&	1166396	\\
\hline
\hline
\end{tabular}

\newpage
\captionof{table}{Listing of coordinates, profile type, redshift, effective radius, S\'ersic-index, colours and NYU-VAGC-ID of all 172 galaxies in field sample 2.}
\centering
\begin{tabular}{cccccccccccc}
\hline
\hline
\multicolumn{12}{c}{Field sample 2} \\
\hline
Index & RA. [deg] & DEC. [deg] & Type & z & $R_{e}$ [kpc] & $n$ & $(g-r)_{in}$ & $(g-r)_{in,1}$ & $(g-r)_{in,2}$ & $(g-r)_{out}$ & ID\\
\hline
1	&	131.463	&	31.9604	&	I 	&	0.0472	&	2.340	&	1.403	&	0.2391	&	0.2152	&	0.2078	&	-	&	1091789	\\
2	&	166.242	&	6.1212	&	I	&	0.0298	&	2.407	&	1.492	&	0.1312	&	0.1194	&	0.1165	&	-	&	0924271	\\
3	&	158.996	&	34.9529	&	I	&	0.0492	&	2.068	&	1.216	&	0.1839	&	0.1877	&	0.1965	&	-	&	1890683	\\
4	&	200.407	&	47.7353	&	I	&	0.0620	&	3.788	&	1.040	&	0.2832	&	0.2723	&	0.2641	&	-	&	1208166	\\
5	&	199.225	&	47.5285	&	I	&	0.0618	&	1.027	&	2.295	&	0.3528	&	0.3491	&	0.3453	&	-	&	1177594	\\
6	&	232.567	&	35.4766	&	I	&	0.0605	&	2.802	&	1.701	&	0.1736	&	0.1467	&	0.1206	&	-	&	1393032	\\
\hline
\hline
\end{tabular}
\newpage
\centering

\begin{tabular}{cccccccccccc}
\hline
\hline
\multicolumn{12}{c}{Field sample 2 (continued)} \\
\hline
Index & RA. [deg] & DEC. [deg] & Type & z & $R_{e}$ [kpc] & $n$ & $(g-r)_{in}$ & $(g-r)_{in,1}$ & $(g-r)_{in,2}$ & $(g-r)_{out}$ & ID\\
\hline
60	&	243.559	&	32.7621	&	II	&	0.0621	&	4.821	&	1.422	&	0.2663	&	0.2373	&	0.2050	&	0.2335	&	1414848	\\
61	&	199.574	&	13.4167	&	II	&	0.0498	&	2.909	&	1.439	&	0.2522	&	0.2215	&	0.2073	&	0.2049	&	1415975	\\
62	&	201.786	&	28.2623	&	II	&	0.0369	&	2.867	&	1.732	&	0.0466	&	0.0115	&	-0.0135	&	-0.0852	&	1977874	\\
63	&	154.050	&	3.6506	&	II	&	0.0320	&	2.089	&	1.645	&	0.2535	&	0.2378	&	0.2361	&	0.2864	&	0445454	\\
64	&	154.318	&	4.7619	&	II	&	0.0623	&	3.839	&	1.353	&	0.3548	&	0.3301	&	0.3147	&	0.3517	&	0450405	\\
65	&	236.194	&	22.8382	&	II	&	0.0452	&	4.033	&	1.531	&	0.3497	&	0.3147	&	0.2992	&	0.2881	&	1967239	\\
\hline
\hline
\end{tabular}
\newpage

\begin{tabular}{cccccccccccc}
\hline
\hline
\multicolumn{12}{c}{Field sample 2 (continued)} \\
\hline
Index & RA. [deg] & DEC. [deg] & Type & z & $R_{e}$ [kpc] & $n$ & $(g-r)_{in}$ & $(g-r)_{in,1}$ & $(g-r)_{in,2}$ & $(g-r)_{out}$ & ID\\
\hline
119	&	253.295	&	40.6836	&	II	&	0.0614	&	4.831	&	1.014	&	0.2438	&	0.2066	&	0.1862	&	0.2005	&	0584808	\\
120	&	231.638	&	4.7850	&	II	&	0.0416	&	1.764	&	1.993	&	0.4663	&	0.4444	&	0.4290	&	0.5231	&	0599710	\\
121	&	202.718	&	58.3555	&	II	&	0.0592	&	1.437	&	1.889	&	0.5591	&	0.5437	&	0.5274	&	0.5436	&	0512268	\\
122	&	241.187	&	49.4584	&	II	&	0.0242	&	3.604	&	1.848	&	0.3199	&	0.2948	&	0.2883	&	0.2777	&	0530131	\\
123	&	125.054	&	18.0934	&	III	&	0.0610	&	3.158	&	1.311	&	0.4732	&	0.4850	&	0.5016	&	0.5017	&	2163133	\\
124	&	188.588	&	29.0042	&	III	&	0.0607	&	1.297	&	2.292	&	0.3951	&	0.3857	&	0.3704	&	0.3837	&	2199946	\\
\hline
\hline
\end{tabular}
\newpage
\captionof{table}{Listing of coordinates, profile type, redshift, effective radius, S\'ersic-index, colours and NYU-VAGC-ID of all 177 galaxies in field sample 3.}
\centering
\begin{tabular}{cccccccccccc}
\hline
\hline
\multicolumn{12}{c}{Field sample 3} \\
\hline
Index & RA. [deg] & DEC. [deg] & Type & z & $R_{e}$ [kpc] & $n$ & $(g-r)_{in}$ & $(g-r)_{in,1}$ & $(g-r)_{in,2}$ & $(g-r)_{out}$ & ID\\
\hline
1	&	224.994	&	39.1991	&	I	&	0.0344	&	1.667	&	1.329	&	0.2237	&	0.2237	&	0.2237	&	-	&	1204128	\\
2	&	248.761	&	26.1807	&	I	&	0.0424	&	2.312	&	1.726	&	0.1430	&	0.1430	&	0.1430	&	-	&	1213783	\\
3	&	167.723	&	10.2944	&	I	&	0.0463	&	2.488	&	1.297	&	0.2084	&	0.2084	&	0.2084	&	-	&	1061491	\\
4	&	236.623	&	11.1031	&	I	&	0.0470	&	1.826	&	0.955	&	0.3308	&	0.3308	&	0.3308	&	-	&	2384542	\\
5	&	141.026	&	51.7645	&	I	&	0.0471	&	2.380	&	2.331	&	0.3982	&	0.3982	&	0.3982	&	-	&	0472596	\\
6	&	131.908	&	6.5121	&	I	&	0.0510	&	2.209	&	1.536	&	0.2808	&	0.2646	&	0.2533	&	-	&	0945534	\\
\hline
\hline
\end{tabular}
\newpage
\centering

\begin{tabular}{cccccccccccc}
\hline
\hline
\multicolumn{12}{c}{Field sample 3 (continued)} \\
\hline
Index & RA. [deg] & DEC. [deg] & Type & z & $R_{e}$ [kpc] & $n$ & $(g-r)_{in}$ & $(g-r)_{in,1}$ & $(g-r)_{in,2}$ & $(g-r)_{out}$ & ID\\
\hline
60	&	199.344	&	7.3986	&	II	&	0.0495	&	4.588	&	1.864	&	0.0372	&	0.0124	&	0.0042	&	0.1261	&	1284407	\\
61	&	135.573	&	33.6371	&	II	&	0.0504	&	2.529	&	1.532	&	0.2034	&	0.1687	&	0.1521	&	-0.0373	&	1881189	\\
62	&	249.541	&	28.7561	&	II	&	0.0543	&	5.852	&	1.446	&	0.2426	&	0.2238	&	0.2054	&	0.1973	&	1415375	\\
63	&	187.759	&	40.2167	&	II	&	0.0570	&	3.007	&	2.176	&	0.2867	&	0.2338	&	0.1959	&	-0.2567	&	1856592	\\
64	&	241.911	&	34.9636	&	II	&	0.0532	&	3.758	&	1.726	&	0.2665	&	0.2286	&	0.2080	&	0.1406	&	1408363	\\
65	&	221.547	&	11.8652	&	II	&	0.0576	&	3.254	&	1.137	&	0.1626	&	0.1309	&	0.1124	&	0.1760	&	1419924	\\
\hline
\hline
\end{tabular}
\newpage

\begin{tabular}{cccccccccccc}
\hline
\hline
\multicolumn{12}{c}{Field sample 3 (continued)} \\
\hline
Index & RA. [deg] & DEC. [deg] & Type & z & $R_{e}$ [kpc] & $n$ & $(g-r)_{in}$ & $(g-r)_{in,1}$ & $(g-r)_{in,2}$ & $(g-r)_{out}$ & ID \\
\hline
119	&	224.239	&	61.8820	&	II	&	0.0470	&	3.346	&	1.414	&	0.2648	&	0.2225	&	0.2016	&	0.1018	&	0501596	\\
120	&	249.617	&	45.4970	&	II	&	0.0583	&	4.020	&	1.560	&	0.3369	&	0.3003	&	0.2884	&	0.2636	&	0505042	\\
121	&	189.959	&	3.7970	&	II	&	0.0400	&	3.468	&	1.277	&	0.0177	&	-0.0009	&	-0.0139	&	-0.0649	&	0520516	\\
122	&	219.074	&	5.5152	&	II	&	0.0601	&	1.912	&	2.216	&	0.5861	&	0.5779	&	0.5707	&	0.5243	&	0597873	\\
123	&	225.284	&	25.4488	&	II	&	0.0597	&	3.811	&	1.903	&	0.1554	&	0.1276	&	0.1083	&	0.1648	&	1933545	\\
124	&	137.497	&	2.7335	&	II	&	0.0378	&	1.777	&	2.181	&	0.3386	&	0.3051	&	0.2786	&	0.2963	&	0444943	\\
\hline
\hline
\end{tabular}
\newpage
\captionof{table}{Listing of scale lengths and break radii (each in both observed bands) of all 175 galaxies in the total cluster sample.}
\centering
\begin{tabular}{ccccccc|ccccccc}
\hline
\hline
\multicolumn{14}{c}{Total cluster sample} \\
\hline
Index & $h_{1,g}$ & $h_{2,g}$ & $R_{b,g}$ & $h_{1,r}$ & $h_{2,r}$ & $R_{b,r}$ & Index & $h_{1,g}$ & $h_{2,g}$ & $R_{b,g}$ & $h_{1,r}$ & $h_{2,r}$ & $R_{b,r}$ \\
& [kpc] & [kpc] & [kpc] & [kpc] & [kpc] & [kpc] & & [kpc] & [kpc] & [kpc] & [kpc] & [kpc] & [kpc] \\
\hline
1 & 1.522 & - & - & 1.5918 & - & - & 60 & 3.4941 & 1.7323 & 5.5227 & 4.4538 & 1.5988 & 5.5023 \\
2 & 2.0167 & - & - & 1.9513 & - & - & 61 & 3.6215 & 1.4766 & 4.0947 & 4.5632 & 1.0642 & 4.5005 \\
3 & 2.6589 & - & - & 2.6359 & - & - & 62 & 4.388 & 2.1286 & 12.5354 & 3.8861 & 2.2727 & 13.0663 \\
4 & 2.0807 & - & - & 2.0176 & - & - & 63 & 5.7654 & 0.9924 & 9.7619 & 7.5603 & 1.2364 & 8.0308 \\
5 & 0.8817 & - & - & 0.9899 & - & - & 64 & 2.7129 & 1.4876 & 6.3109 & 2.5075 & 1.4646 & 7.2546 \\
\hline
\hline
\end{tabular}
\newpage

\begin{tabular}{ccccccc|ccccccc}
\hline
\hline
\multicolumn{14}{c}{Total cluster sample (continued)} \\
\hline
Index & $h_{1,g}$ & $h_{2,g}$ & $R_{b,g}$ & $h_{1,r}$ & $h_{2,r}$ & $R_{b,r}$ & Index & $h_{1,g}$ & $h_{2,g}$ & $R_{b,g}$ & $h_{1,r}$ & $h_{2,r}$ & $R_{b,r}$ \\
& [kpc] & [kpc] & [kpc] & [kpc] & [kpc] & [kpc] & & [kpc] & [kpc] & [kpc] & [kpc] & [kpc] & [kpc] \\
\hline
119 & 4.7734 & 1.4565 & 2.7753 & 2.912 & 1.4069 & 3.1332 & 148 & 1.2165 & 1.5618 & 5.0213 & 1.2598 & 1.5259 & 5.8237 \\
120 & 4.8657 & 2.1717 & 4.2166 & 3.5741 & 2.0416 & 4.8791 & 149 & 1.659 & 2.8103 & 4.0826 & 1.3541 & 2.6561 & 3.649 \\
121 & 1.1803 & 1.1027 & 4.1682 & 1.238 & 0.9634 & 4.2718 & 150 & 0.8279 & 0.9781 & 2.4463 & 0.8465 & 1.0728 & 2.7055 \\
122 & 1.6898 & 0.3969 & 6.5095 & 1.6462 & 0.7345 & 5.9498 & 151 & 1.1643 & 2.5775 & 3.5959 & 1.0886 & 2.422 & 3.4859 \\
123 & 6.9686 & 2.0607 & 4.2102 & 3.6438 & 2.1453 & 4.8422 & 152 & 1.0131 & 4.4978 & 4.7638 & 1.0525 & 2.5445 & 4.5075 \\
\hline
\hline
\end{tabular}
\vspace{0.25cm}
\captionof{table}{Listing of scale lengths and break radii (each in both observed bands) of all 172 galaxies in field sample 1.}
\begin{tabular}{ccccccc|ccccccc}
\hline
\hline
\multicolumn{14}{c}{Field sample 1} \\
\hline
Index & $h_{1,g}$ & $h_{2,g}$ & $R_{b,g}$ & $h_{1,r}$ & $h_{2,r}$ & $R_{b,r}$ & Index & $h_{1,g}$ & $h_{2,g}$ & $R_{b,g}$ & $h_{1,r}$ & $h_{2,r}$ & $R_{b,r}$ \\
& [kpc] & [kpc] & [kpc] & [kpc] & [kpc] & [kpc] & & [kpc] & [kpc] & [kpc] & [kpc] & [kpc] & [kpc] \\
\hline
1 & 0.8365 & - & - & 0.8815 & - & - & 29 & 2.4166 & 0.8585 & 4.1297 & 2.4313 & 0.8964 & 4.4879 \\
2 & 2.0981 & - & - & 2.2205 & - & - & 30 & 1.9146 & 1.4031 & 7.4893 & 2.157 & 1.3332 & 8.1887 \\
3 & 1.434 & - & - & 1.6664 & - & - & 31 & 3.1575 & 1.5007 & 6.2759 & 2.9816 & 1.3807 & 5.8988 \\
4 & 1.4207 & - & - & 1.4155 & - & - & 32 & 1.0236 & 0.2849 & 5.4294 & 1.7018 & 0.6525 & 3.9984 \\
5 & 0.8399 & - & - & 0.875 & - & - & 33 & 3.1062 & 1.4664 & 5.5548 & 2.2586 & 1.213 & 7.4191 \\
\hline
\hline
\end{tabular}

\newpage
\begin{tabular}{ccccccc|ccccccc}
\hline
\hline
\multicolumn{14}{c}{Field sample 1 (continued)} \\
\hline
Index & $h_{1,g}$ & $h_{2,g}$ & $R_{b,g}$ & $h_{1,r}$ & $h_{2,r}$ & $R_{b,r}$ & Index & $h_{1,g}$ & $h_{2,g}$ & $R_{b,g}$ & $h_{1,r}$ & $h_{2,r}$ & $R_{b,r}$ \\
& [kpc] & [kpc] & [kpc] & [kpc] & [kpc] & [kpc] & & [kpc] & [kpc] & [kpc] & [kpc] & [kpc] & [kpc] \\
\hline
57 & 1.9983 & 0.7781 & 2.7765 & 1.6055 & 1.0027 & 2.7766 & 115 & 2.2146 & 1.4492 & 6.4415 & 2.4739 & 1.3608 & 7.1646 \\
58 & 4.1316 & 1.3267 & 4.1026 & 4.5374 & 1.3233 & 3.9627 & 116 & 6.3023 & 1.9008 & 4.9987 & 4.7875 & 2.0203 & 5.0497 \\
59 & 1.8094 & 0.9097 & 3.6256 & 1.9821 & 0.998 & 3.7139 & 117 & 1.686 & 1.2626 & 4.8524 & 1.8433 & 1.3158 & 5.0365 \\
60 & 2.8628 & 1.6007 & 7.3947 & 3.3724 & 1.7701 & 7.2711 & 118 & 2.8235 & 2.0992 & 2.9484 & 2.6263 & 1.9455 & 3.3173 \\
61 & 1.6009 & 1.1951 & 3.8489 & 1.771 & 1.1583 & 4.692 & 119 & 3.5053 & 1.6007 & 8.3146 & 3.5308 & 1.8005 & 8.2778 \\
\hline
\hline
\end{tabular}
\newpage
\captionof{table}{Listing of scale lengths and break radii (each in both observed bands) of all 172 galaxies in field sample 2.}
\begin{tabular}{ccccccc|ccccccc}
\hline
\hline
\multicolumn{14}{c}{Field sample 2} \\
\hline
Index & $h_{1,g}$ & $h_{2,g}$ & $R_{b,g}$ & $h_{1,r}$ & $h_{2,r}$ & $R_{b,r}$ & Index & $h_{1,g}$ & $h_{2,g}$ & $R_{b,g}$ & $h_{1,r}$ & $h_{2,r}$ & $R_{b,r}$ \\
& [kpc] & [kpc] & [kpc] & [kpc] & [kpc] & [kpc] & & [kpc] & [kpc] & [kpc] & [kpc] & [kpc] & [kpc] \\
\hline
1 & 1.4816 & - & - & 1.5247 & - & - & 60 & 5.5352 & 2.5674 & 4.9389 & 3.4597 & 1.9148 & 7.835 \\
2 & 1.2405 & - & - & 1.2534 & - & - & 61 & 2.3655 & 1.6031 & 2.7878 & 2.6825 & 1.4852 & 2.6287 \\
3 & 1.465 & - & - & 1.3931 & - & - & 62 & 3.6894 & 1.528 & 4.1191 & 3.0415 & 1.3483 & 3.9019 \\
4 & 1.784 & - & - & 1.7463 & - & - & 63 & 2.2893 & 1.1203 & 2.6989 & 2.3261 & 0.9516 & 2.6178 \\
5 & 1.016 & - & - & 0.9875 & - & - & 64 & 3.1491 & 1.8965 & 4.8273 & 2.7258 & 1.8116 & 4.9298 \\
\hline
\hline
\end{tabular}
\newpage
\begin{tabular}{ccccccc|ccccccc}
\hline
\hline
\multicolumn{14}{c}{Field sample 2 (continued)} \\
\hline
Index & $h_{1,g}$ & $h_{2,g}$ & $R_{b,g}$ & $h_{1,r}$ & $h_{2,r}$ & $R_{b,r}$ & Index & $h_{1,g}$ & $h_{2,g}$ & $R_{b,g}$ & $h_{1,r}$ & $h_{2,r}$ & $R_{b,r}$ \\
& [kpc] & [kpc] & [kpc] & [kpc] & [kpc] & [kpc] & & [kpc] & [kpc] & [kpc] & [kpc] & [kpc] & [kpc] \\
\hline
119 & 7.2262 & 1.9807 & 5.7089 & 5.5944 & 1.9109 & 5.5108 & 146 & 1.4859 & 1.8961 & 3.4409 & 1.1042 & 1.4324 & 5.2665 \\
120 & 2.2247 & 0.9597 & 4.2305 & 1.886 & 0.6792 & 4.8339 & 147 & 1.5657 & 1.7075 & 3.0479 & 1.3783 & 1.5445 & 3.7113 \\
121 & 1.5644 & 0.9712 & 2.5997 & 1.5508 & 0.6807 & 4.3251 & 148 & 1.1417 & 1.1729 & 1.5143 & 0.8994 & 1.3434 & 2.7575 \\
122 & 2.05 & 1.7764 & 4.9833 & 1.9776 & 1.5395 & 5.1326 & 149 & 1.6074 & 1.8491 & 4.7117 & 1.3344 & 1.8995 & 4.9746 \\
123 & 1.9191 & 1.8852 & 4.5002 & 1.8785 & 1.8328 & 5.4535 & 150 & 1.0203 & 0.9059 & 1.2367 & 0.7902 & 0.8086 & 1.1806 \\
\hline
\hline
\end{tabular}

\vspace{0.25cm}
\captionof{table}{Listing of scale lengths and break radii (each in both observed bands) of all 177 galaxies in field sample 3.}
\begin{tabular}{ccccccc|ccccccc}
\hline
\hline
\multicolumn{14}{c}{Field sample 3} \\
\hline
Index & $h_{1,g}$ & $h_{2,g}$ & $R_{b,g}$ & $h_{1,r}$ & $h_{2,r}$ & $R_{b,r}$ & Index & $h_{1,g}$ & $h_{2,g}$ & $R_{b,g}$ & $h_{1,r}$ & $h_{2,r}$ & $R_{b,r}$ \\
& [kpc] & [kpc] & [kpc] & [kpc] & [kpc] & [kpc] & & [kpc] & [kpc] & [kpc] & [kpc] & [kpc] & [kpc] \\
\hline
1 & 1.0678 & - & - & 1.1306 & - & - & 30 & 1.6276 & 1.2349 & 4.0304 & 1.45 & 1.222 & 4.2891 \\
2 & 1.4189 & - & - & 1.4475 & - & - & 31 & 2.3017 & 1.3873 & 3.6776 & 2.5188 & 1.1845 & 3.9571 \\
3 & 1.4982 & - & - & 1.7333 & - & - & 32 & 3.3937 & 1.2106 & 3.9901 & 2.7085 & 1.3278 & 4.0236 \\
4 & 0.9593 & - & - & 0.9772 & - & - & 33 & 2.717 & 2.3257 & 4.2151 & 2.749 & 2.4516 & 3.76 \\
5 & 1.524 & - & - & 1.4227 & - & - & 34 & 1.0647 & 0.7184 & 5.3328 & 0.9814 & 0.9368 & 7.4206 \\
\hline
\hline
\end{tabular}

\newpage

\begin{tabular}{ccccccc|ccccccc}
\hline
\hline
\multicolumn{14}{c}{Field sample 3 (continued)} \\
\hline
Index & $h_{1,g}$ & $h_{2,g}$ & $R_{b,g}$ & $h_{1,r}$ & $h_{2,r}$ & $R_{b,r}$ & Index & $h_{1,g}$ & $h_{2,g}$ & $R_{b,g}$ & $h_{1,r}$ & $h_{2,r}$ & $R_{b,r}$ \\
& [kpc] & [kpc] & [kpc] & [kpc] & [kpc] & [kpc] & & [kpc] & [kpc] & [kpc] & [kpc] & [kpc] & [kpc] \\
\hline
59 & 1.3917 & 1.018 & 5.3009 & 3.4525 & 0.7782 & 4.5677 & 119 & 2.6521 & 2.0144 & 4.9589 & 2.2288 & 1.9407 & 5.4681 \\
60 & 3.7486 & 0.9864 & 11.4976 & 3.5611 & 1.2757 & 11.6012 & 120 & 3.0175 & 2.4968 & 5.2479 & 2.6647 & 2.3622 & 4.281 \\
61 & 1.9369 & 0.9868 & 5.5094 & 1.7398 & 0.403 & 7.0213 & 121 & 2.7888 & 2.1841 & 9.0678 & 2.5945 & 2.4012 & 9.0154 \\
62 & 4.3387 & 3.5488 & 8.5731 & 3.8914 & 3.6311 & 9.1993 & 122 & 2.0951 & 1.5514 & 5.5556 & 1.9791 & 1.6738 & 5.1832 \\
63 & 3.2221 & 2.2144 & 12.3918 & 2.9778 & 1.4124 & 12.7983 & 123 & 4.5069 & 2.2691 & 3.7368 & 5.3201 & 2.7862 & 4.0238 \\
\hline
\hline
\end{tabular}
\label{lastpage}

\end{document}